\documentclass[aps,pra,reprint,groupedaddress,showpacs]{revtex4-1}
\usepackage{graphicx}
\usepackage{latexsym,amssymb,amsmath}
\usepackage{bm}
\usepackage[caption=false]{subfig}
\usepackage{xcolor}
\usepackage{mathrsfs}
\usepackage{hyperref}

\begin{document}
\title{Achieving strongly negative scattering asymmetry factor in random media composed of dual-dipolar particles}
\author{B. X. Wang}
\affiliation{Institute of Engineering Thermophysics, Shanghai Jiao Tong University, Shanghai, 200240, P. R. China}
\author{C. Y. Zhao}
\email{Changying.zhao@sjtu.edu.cn}
\affiliation{Institute of Engineering Thermophysics, Shanghai Jiao Tong University, Shanghai, 200240, P. R. China}
\date{\today}
\begin{abstract}
Understanding radiative transfer in random media like micro/nanoporous and particulate materials, allows people to manipulate the scattering and absorption of radiation, as well as opens new possibilities in applications such as imaging through turbid media, photovoltaics and radiative cooling. A strong-backscattering phase function, i.e., a negative scattering asymmetry parameter $g$, is of great interest, which can possibly lead to unusual radiative transport phenomena, for instance, Anderson localization of light. Here we demonstrate that by utilizing the structural correlations and second Kerker condition for a disordered medium composed of randomly distributed silicon nanoparticles, a strongly negative scattering asymmetry factor ($g\sim-0.5$) for multiple light scattering can be realized in the near-infrared. Based on the multipole expansion of Foldy-Lax equations and quasicrystalline approximation (QCA), we have rigorously derived analytical expressions for the effective propagation constant and scattering phase function for a random system containing spherical particles, by taking the effect of structural correlations into account. We show that as the concentration of scattering particles rises, the backscattering is also enhanced. Moreover, in this circumstance, the transport mean free path is largely reduced and even becomes smaller than that predicted by independent scattering approximation. We further explore the dependent scattering effects, including the modification of electric and magnetic dipole excitations and far-field interference effect, both induced and influenced by the structural correlations, for volume fraction of particles up to $f_v\sim0.25$. Our results have profound implications in harnessing micro/nanoscale radiative transfer through random media.  
\end{abstract}

%\keywords{}

\maketitle

\section{Introduction}
Rich interference phenomena in random or disordered media have received growing attention in the last a few years, and give rise to a rapidly developing field called ``disordered photonics'' \cite{wiersma2013disordered,Rotter2017}. The study of disordered photonics involves the fundamental pursuit of Anderson localization of light in various disordered micro/nanostructures \cite{wiersma1997localization,Storzer2006,Segev2013,Sperling2016NJP}, random lasers \cite{Cao1999,Wiersma2008}, amorphous photonic crystals \cite{Florescu2009,Froufe-Perez2016}, photovoltaics \cite{Vynck2012,Fang2015JQSRT,Liew2016ACSPh}, focusing \cite{Vellekoop2010} and imaging \cite{Mosk2012,hsu2017correlation}, etc. When light propagates in disordered photonic media, it undergoes scattering in a very complicated way. Traditionally, the transport of light intensity is depicted by the radiative transfer equation (RTE) \cite{lagendijk1996resonant,VanRossum1998,tsang2004scattering,sheng2006introduction,mishchenko2006multiple,akkermans2007mesoscopic,mishchenko2014electromagnetic}. Parameters describing radiative transport, particularly the scattering mean free path $l_\text{s}$ and the transport mean free path $l_\text{tr}$, are usually calculated under the independent scattering approximation (ISA), i.e., in which the scatterers scatter electromagnetic waves independently without any inter-particle interference taken into account \cite{lagendijk1996resonant,VanRossum1998,tsang2004scattering,sheng2006introduction,akkermans2007mesoscopic}. $l_\text{s}$ and $l_\text{tr}$ are related through the single particle scattering asymmetry factor $g$, i.e., the mean cosine of the scattering angle of the scattering phase function, as $l_\text{tr}=l_\text{s}/(1-g)$. Since $g\ge0$ is valid for most natural scatterers, it is common to conclude that $l_\text{tr}\geq l_\text{s}$. 

For diffusive light transport in three-dimensional random media (the length scale along the propagation direction $L\gg l_\text{tr}$), the diffusion constant $D$ is related to the transport mean free path via $D=v_\text{E}l_\text{tr}/3$, where $v_\text{E}$ is the energy transport velocity in random media \cite{sheng2006introduction}. Moreover, the Ioffe-Regel parameter $k_\text{e}l_\text{tr}$, which decides whether the scattering strength is strong enough to make Anderson localization occur, also depends on $l_\text{tr}$, where $k_\text{e}=2\pi/\lambda_\text{e}$ is the wavevector and $\lambda_\text{e}$ is the renormalized wavelength in the random media \cite{sheng2006introduction}. Therefore, it is critical to lower $l_\text{tr}$ to achieve strong light scattering and thus mediate a transition to Anderson localization as well as other unusual transport phenomena \cite{sheng2006introduction,akkermans2007mesoscopic}. A possible way is to artificially create negative asymmetry factor $g$ to make $l_\text{tr}<l_\text{s}$. Decades ago, it was already reported by Pinhero et al. that magnetic particles with giant permeability $\mu\gg1$ can induce negative $g$ in the visible range \cite{pinheroPRL2000}. Even more earlier, in the 1980s Kerker et al. \cite{kerkerJOSA1983} already proposed critical conditions in which perfect directional scattering for single magneto-dielectric particles can be realized. Specifically, by properly tuning the amplitude and spectral position of electric and magnetic dipoles, one can obtain a nearly-zero-froward scattering (NZFS) pattern for these particles. This is called the second Kerker condition, where the permittivity $\varepsilon$ and permeability $\mu$ of a very small particle should satisfy the condition $\varepsilon=(4-\mu)/(2\mu+1)$ with $\mu\neq1$. This feature arises from the destructive interference of electric and magnetic dipoles in the forward direction. Note the original proposal of zero forward scattering condition of Kerker et al. violates the optical theorem, and thus the term ``nearly-ZFS'' is used here, which indicates the forward scattering amplitude is not rigorously zero \cite{naraghiOL2015}.

Later, many researchers showed that actually some nonmagnetic particles with moderate permittivity can also exhibit NZFS feature if one appropriately excites the electric and magnetic dipoles in the particle. For instance, recent advances in all dielectric metasurfaces revealed that by carefully modulating the sizes of moderate-refractive-index dielectric nanoparticles, spectral overlapping of electric and magnetic dipoles can be achieved, resulting in NZFS, e.g., for silicon \cite{geffrinNC2012,Gomez-MedinaPRA2012} or germanium \cite{Gomez-MedinaJNP2011} nanospheres. Actually when the first-order Mie coefficients $a_1$ (describing electric diple response) and $b_1$ (describing magnetic dipole response) have the same amplitude and are out of phase with each other, i.e., $a_1=-b_1$, the second Kerker condition is perfectly satisfied and the particle shows a NZFS scattering pattern. In particular, G\'{o}mez-Medina et al. \cite{Gomez-MedinaPRA2012} found that for a single silicon particle with radius $a=230\mathrm{nm}$, a negative asymmetry factor as small as $g=-0.15$ at $\lambda=1530\mathrm{nm}$ can be obtained. Note in these cases high-order Mie multipolar modes in the particles are negligible, which can be then termed ``dual-dipolar particles''\cite{Zambrana-PuyaltoOL2013,Zambrana-PuyaltoOE2013,schmidtPRL2015}.
 
Above discussions are specific for the single particle scattering regime, only valid for random media containing very dilutely distributed particles \cite{VanRossum1998}. As the concentration of scattering particles rises, ISA becomes unsuitable due to the existence of interparticle correlations \cite{fradenPRL1990,mishchenkoJQSRT1994}, which make the scattered electromagnetic waves of different particles interference and preserve partial coherence \cite{laxRMP1951}. In this case, the scattering asymmetry factor $g$ for the whole random media is rather different from that of single scattering \cite{fradenPRL1990,mishchenkoJQSRT1994,mishchenkoJQSRT1997,mishchenkoKPCB2010}. The influence of the structural correlations actually provides an alternative way to control the asymmetry factor of light transport in random media, as was done by Rojas-Ochoa et al.\cite{rojasochoaPRL2004}, who used dense colloidal suspensions of polystyrene with inter-particle repulsive electrostatic forces to achieve a very negative asymmetry factor around $g\sim-1$. They showed that it is the structural short-range-order induced Bragg backscattering resonance that leads to this extremely negative $g$. 

In this study, by cooperatively utilizing the common hard-sphere structural correlations and the second Kerker condition for dual-dipolar particles, we demonstrate that strong backscattering (negative asymmetry factor $g\sim-0.5$) can still be achieved, not relying on local Bragg resonance. By means of the multipole expansion method and quasicrystalline approximation (QCA) for the Foldy-Lax equations (FLEs) treating multiple scattering of electromagnetic waves, we rigorously derive analytical expressions for the effective propagation constant and scattering phase function for the random system consisting of dual-dipolar particles. The obtained transport mean free path $l_\text{tr}$ is shown to be substantially shorter than the scattering mean free path $l_\text{s}$. We also address the dependent scattering mechanism and its interplay with the structural correlations, or short-range order, which then allows a flexible control over light-matter interaction in random media. The negative $g$ leads to an unusual multiple scattering regime, implying a possible way for realizing three-dimensional Anderson localization and other anomalous transport phenomena. It is promising to utilize negative asymmetry factor to achieve extreme light-matter interaction, facilitating the performance of novel photonic devices like random lasers, disordered photonic bandgap media, and light trapping and conversion devices, etc.
\section{Negative asymmetry factor for a single particle}\label{single_particle}
The scattering of electromagnetic waves by single homogeneous or multilayered spherical particles with arbitrary electric and magnetic properties is one of the earliest solved problems in electromagnetic scattering, which was done by Gustav Mie over 100 years ago \cite{mie1908}. Along with the rapid development of nanofabrication and nanophotonics in the last a few years, the anomalous scattering properties of single dielectric particles are theoretically and experimentally studied by many authors very extensively \cite{tribelskyPRL2006,tribelskyEPL2011,kuznetsovSR2012,fuNC2013,kuznetsovScience2016,Luk'yanchukPRA2017,valuckasAPL2017}, giving rise to the booming of nanoscale light scattering study. The basic idea behind Mie theory is to rigorously solve the boundary value problem of Maxwell's equations in spherical coordinates. For spherical boundary conditions, the solution of Maxwell's equations can be formally expanded in a linear combination of vector spherical harmonics (VSHs) or vector spherical wave functions (VSWFs) \cite{bohrenandhuffman,tsang2000scattering1}. The extinction and scattering efficiencies of a single homogeneous sphere with a complex refractive index of $\tilde{m}$ and radius $a$ placed in vacuum illuminated by a plane wave is formally given by \cite{Gomez-MedinaPRA2012,bohrenandhuffman,tsang2000scattering1}
\begin{equation}\label{cext_eq}
C_{\text{ext}}=\frac{2\pi}{k^2}\sum_{n=1}^{\infty}(2n+1)\mathrm{Re}(a_n+b_n),
\end{equation}
\begin{equation}
C_{\text{sca}}=\frac{2\pi}{k^2}\sum_{n=1}^{\infty}(2n+1)(|a_n|^2+|b_n|^2),
\end{equation}
where 
\begin{equation}
a_n=\frac{\tilde{m}^2j_n(\tilde{m}x)[xj_n(x)]'-j_n(x)[\tilde{m}xj_n(\tilde{m}x)]'}{\tilde{m}^2j_n(\tilde{m}x)[xh_n(x)]'-h_n(x)[\tilde{m}xj_n(\tilde{m}x)]'},
\end{equation}
\begin{equation}
b_n=\frac{j_n(\tilde{m}x)[xj_n(x)]'-j_n(x)[\tilde{m}xj_n(\tilde{m}x)]'}{j_n(\tilde{m}x)[xh_n(x)]'-h_n(x)[\tilde{m}xj_n(\tilde{m}x)]'},
\end{equation}
and $k=2\pi/\lambda$ is the wavenumber of plane wave with wavelength $\lambda$ and $x=ka$ is the corresponding size parameter. $j_n(z)$  and $h_n(z)$ are spherical Bessel functions and Hankel functions of the first kind of order $n$, with respect to the argument $z$ \cite{bohrenandhuffman}.
\begin{figure}[htbp]
	\centering
	\subfloat{
		\label{figqext}
		\includegraphics[width=0.8\linewidth]{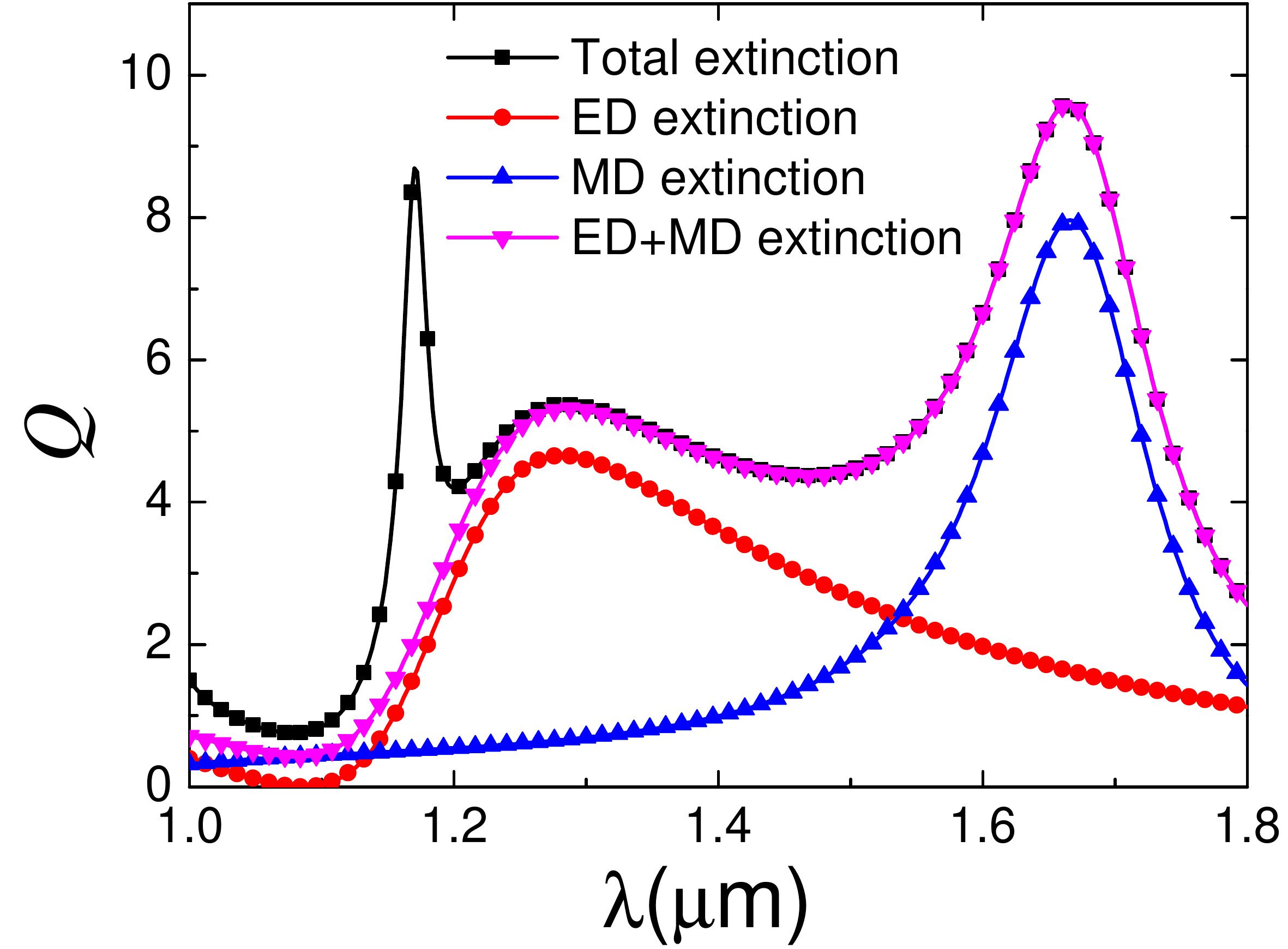}
	}
	\hfill
    \subfloat{
    	\label{figg1}
    	\includegraphics[width=0.8\linewidth]{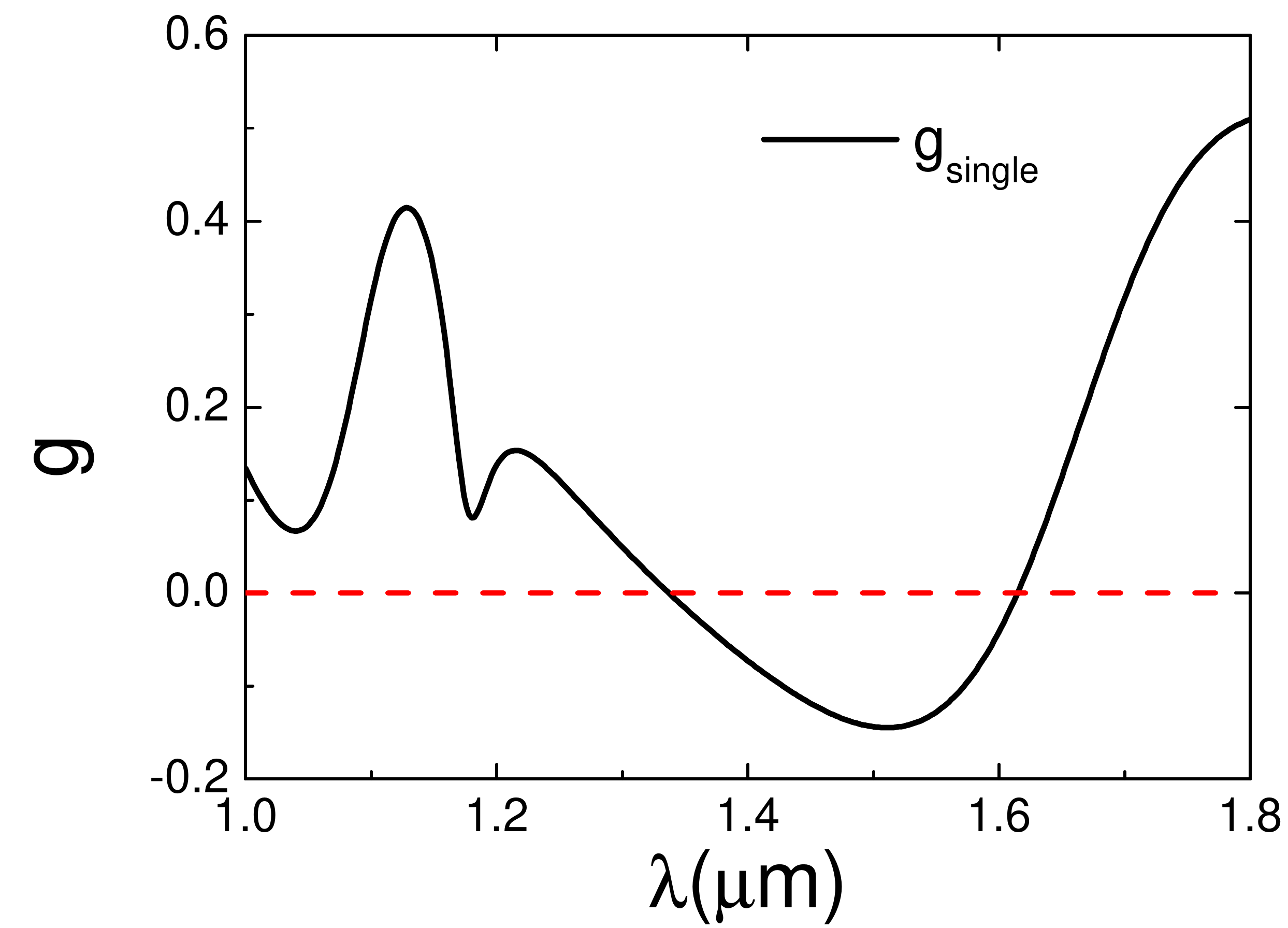}
    }
	\caption{Extinction efficiency for a single Si nanoparticle with radius $a=230\mathrm{nm}$, where the contributions of electric dipole (ED) and magnetic dipole (MD) as well as the sum of them (ED+MD) are also shown.}
	
\end{figure} 
%\begin{figure}[htbp]
%	\centering
%	\includegraphics[width=0.8\linewidth]{g1}
%	\caption{Scattering asymmetry factor for a single Si nanoparticle with radius $a=230\mathrm{nm}$ under different wavelength.}
%	\label{figg1}
%\end{figure}

The differential scattering cross section for unpolarized incident light is given by
\begin{equation}\label{pf_isa}
\frac{d\sigma_{\text{s}}}{d\theta_{\text{s}}}=\frac{\pi}{k^2}(|S_1(\theta_{\text{s}})|^2+|S_2(\theta_{\text{s}})|^2),
\end{equation}
where 
\begin{equation}\label{s1_isa}
S_1(\theta_{\text{s}})=\sum_{n=1}^{\infty}\frac{2n+1}{n(n+1)}[a_n\pi_n(\cos{\theta_{\text{s}}})+b_n\tau_n(\cos{\theta_{\text{s}}})]
\end{equation}
and
\begin{equation}\label{s2_isa}
S_2(\theta_{\text{s}})=\sum_{n=1}^{\infty}\frac{2n+1}{n(n+1)}[a_n\tau_n(\cos{\theta_{\text{s}}})+b_n\pi_n(\cos{\theta_{\text{s}}})]
\end{equation}
are elements of amplitude scattering matrix and $\theta_s$ is the polar scattering angle with respect to the incident wavevector, in which $\pi_n$ and $\tau_n$ are functions defined in Appendix \ref{far-field_appendix} \cite{bohrenandhuffman,tsang2000scattering1} . The normalized differential scattering cross section is also called scattering phase function used in RTE. Therefore, the scattering asymmetry factor for the single scattering phase function, defined as the mean cosine of scattering angle, $\langle\cos\theta_{\text{s}}\rangle$, is calculated through \cite{bohrenandhuffman}

\begin{equation}
g=\frac{1}{\sigma_{\text{s}}}\int_{0}^{\pi}\frac{d\sigma_{\text{s}}}{d\theta_{\text{s}}}\cos\theta_{\text{s}}\sin\theta_{\text{s}} d\theta_{\text{s}}.
\end{equation}

Based on Mie theory, it is straightforward to find negative values of $g$ by tuning the material and geometry parameters, as was done by G\'{o}mez-Medina et al. \cite{Gomez-MedinaPRA2012}. Particularly, for a dual-dipolar nanoparticle \cite{schmidtPRL2015}, in which only the electric and magnetic dipole modes are excited, above series sums are reduced to only $n=1$ and simple analytical conditions can be found . This is the case for silicon nanoparticles with radius around $a\sim200\mathrm{nm}$ in the near infrared. For a $a=230\mathrm{nm}$ silicon nanoparticle, the total extinction efficiency $Q_{{\text{ext}}}=C_{\text{ext}}/(\pi a^2)$ is shown in Fig.\ref{figqext}, along with the contribution of electric dipole (ED), magnetic dipole (MD) and their sum (MD+ED) according to Eq. (\ref{cext_eq}). It is clearly seen that for $\lambda\gtrsim1300\mathrm{nm}$, the ED and MD excitations are dominating. In this circumstance, the second Kerker condition requires \cite{naraghiOL2015,Gomez-MedinaPRA2012}
\begin{equation}
|a_1|=|b_1|
\end{equation}
and 
\begin{equation}
|\arg{a_1}-\arg{b_1}|=\pi
\end{equation}
for dual-dipolar particles. A perfect NZFS radiation pattern can be realized if these conditions are fulfilled, leading to the asymmetry factor \cite{Gomez-MedinaPRA2012}
\begin{equation}\label{gdual_dipole_eq}
g_{\text{min}}=\frac{\mathrm{Re}(a_1b_1^*)}{|a_1|^2+|b_1|^2}=-0.5,
\end{equation}
which is the minimum value of $g$ for dual-dipolar particles. For a silicon sphere with $a=230\mathrm{nm}$ shown in Fig. \ref{figqext}, only a moderately negative $g=-0.15$ can be achieved at the equal amplitude wavelength $\lambda=1530\mathrm{nm}$ because $a_1$ and $b_1$ are not ideally out-of-phase. The spectrum of asymmetry factor for this single particle is shown in Fig. \ref{figg1}. This fact leads us to consider whether it is possible to achieve a stronger backscattering by utilizing inteference effects among multiple particles in random media composed of these dual-dipolar particles, especially the interference induced by particle structural correlations. In the next section, we will analytically solve the multiple scattering problem of electromagnetic wave propagation in such random medium to further examine this idea.

\section{Negative asymmetry factor for a random medium composed of  dual-dipolar particles}\label{multi_particle}

\subsection{Effective propagation constant}\label{general_theory}
When a large number of identical dual-dipolar particles approaching the second Kerker condition (Here we choose  Si nanoparticle with radius $a=230\mathrm{nm}$ at $\lambda=1530\mathrm{nm}$.) are randomly packed and constitute a disordered medium, it is of great interest to investigate whether the dependent scattering effect can lead to a more negative asymmetry factor for radiative transfer than the single scattering case. It should be noted here we call the ``dependent scattering effect'' as a generalization for those interference effects that are not possible to explain under ISA \cite{yamadaJHT1986,aernoutsOE2014}. This is a broader definition than that of van Tiggelen et al.'s \cite{Vantiggelen1990JPCM}, for instance, which classified the multiple scattering trajectories visiting the same particle more than once and resulting in a closed loop, or ``recurrent scattering''\cite{Aubry2014PRL}, as the dependent scattering mechanism.  

In this section, we will present an analytical derivation for the dependent scattering effect, starting from the first principles of electromagnetic wave theory. Since the scatterers are randomly distributed in the medium and have finite sizes comparable with the wavelength, it is pivot to take the inter-particle correlations into account in the dependent scattering model \cite{fradenPRL1990,mishchenkoJQSRT1994,rojasochoaPRL2004}. This is because the existence of one particle would create an exclusion volume into which other particles are not allowed to penetrate, which leads to definite phase differences among scattered waves preserving over ensemble average. These definite phase differences produce constructive or destructive interferences which in turn affect the transport properties of light, which are called ``partial coherence'' by Lax \cite{laxRMP1951,laxPR1952}. Therefore, to establish an analytically solvable model, a statistical description of scatterer positions is needed. Typically, the pair distribution function (PDF), $g_2(\mathbf{r}_1,\mathbf{r}_2)$ is used to describe the statistical distribution between a pair of particles, more specifically, the conditional probability density of finding a particle centered at the position $\mathbf{r}_1$ when a fixed particle is seated at $\mathbf{r}_2$.  When assuming the random medium is statistically homogeneous and isotropic, the PDF only depends on the distance between the pair of particles, i.e.,$g_2(\mathbf{r}_1,\mathbf{r}_2)=g_2(|\mathbf{r}_1-\mathbf{r}_2|)$ \cite{wertheimPRL1963,tsang2004scattering2}. There are already several (approximate) analytical solutions of the PDF for some specific random systems, e.g., Refs.\cite{wertheimPRL1963,Baxter1968}. However, analytical expressions for high-order position correlations involving three or more particles simultaneously are much harder to obtain. In this circumstance, if QCA is introduced, these correlations are treated as a hierarchy of PDFs, e.g., $g_3(\mathbf{r}_1,\mathbf{r}_2,\mathbf{r}_3)=g_2(\mathbf{r}_1,\mathbf{r}_2)g_2(\mathbf{r}_2,\mathbf{r}_3)$, $g_4(\mathbf{r}_1,\mathbf{r}_2,\mathbf{r}_3,\mathbf{r}_4)=g_2(\mathbf{r}_1,\mathbf{r}_2)g_2(\mathbf{r}_2,\mathbf{r}_3)g_2(\mathbf{r}_3,\mathbf{r}_4)$ and so forth, where $g_3$ and $g_4$ indicate three-particle and four-particle distribution functions, respectively \cite{laxPR1952,bringi1982coherent,tsangJAP1982,tishkovetsJQSRT2011}. In this one respect, QCA is actually a perturbative approach only incorporating multiple scattering diagrams with two-particle statistics, but it still takes infinite scattering orders into account \cite{maAO1988,varadanJOSAA1985}. This approximation sets up the basis of the present analysis. 

QCA was initially proposed by Lax \cite{laxPR1952} for both quantum and classical waves, and examined by exact numerical simulations as well as experiments to be satisfactorily accurate for the dependent scattering effect in moderately dense random media \cite{westJOSAA1994,Nashashibi1999,Siqueira2000}. It is also widely used in the prediction of optical and radiative properties of disordered materials for applications in remote sensing \cite{liangIEEETGRS2008} as well as thermal radiation transfer \cite{prasherJAP2014,wangIJHMT2015}. More generally, its validity for ultrasonic waves propagation in acoustical random media is also frequently verified numerically and experimentally \cite{meulenJASA2001}. However, there is no rigorous treatment for light propagation, especially for intensity transport, in a random system of dual-dipolar particles in the spirit of QCA \cite{tsang2004scattering}. In the next, by applying the multipole expansion method for the FLEs \cite{laxRMP1951}, QCA and a series of other well-defined approximations, we will rigorously derive analytical expressions for the effective propagation constant and scattering phase function in such medium.
%In diagrammatic representation for the self-energy $\Sigma$,  The diagrammatic representation of QCA for the self-energy of the field is shown in Fig. \ref{figqcadiagram}.
% \begin{figure}[htbp]
% \centering
% \includegraphics[width=0.6\linewidth]{diagram_qca}

\begin{figure}[htbp]
	\centering
	\includegraphics[width=0.8\linewidth]{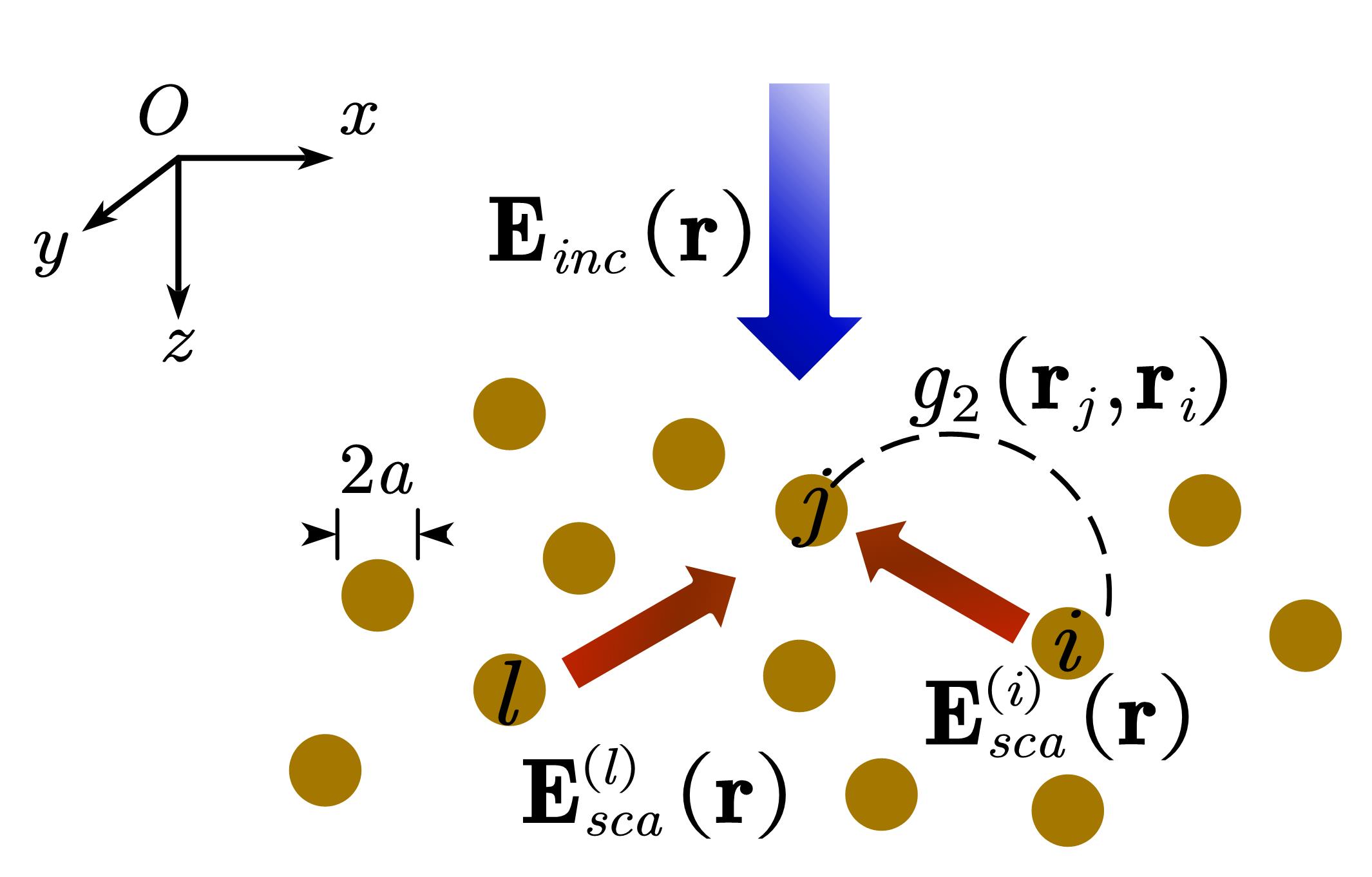}
	\caption{A schematic of Foldy-Lax equations for multiple scattering of electromagnetic waves in randomly distributed spherical particles. The particles are numbered as $i$, $j$, $l$, etc. The dashed line denotes $g_2(\mathbf{r}_j,\mathbf{r}_i)$, the pair distribution function between the two particles. The thick arrow indicates the propagation direction of the incident wave, while the thin arrows stand for the propagation directions of the partial scattered waves from particle $i$ to $j$ and from $l$ to $j$.}
	\label{system_config}
\end{figure}

% \caption{Diagrammatic representation of QCA. Circles denote the T-matrix of single particles, bold solid lines denote free-space propagator and thin solid lines represent particle correlation function $h_2(r)=g_2(r)-1$.}
% \label{figqcadiagram}
% \end{figure}
Before proceeding to our derivation, we stress some basic assumptions about the random medium. The positions of scatterers, for a specific initial configuration of scatterers, are treated to be fixed if they move sufficiently slower than the transport of electromagnetic waves \cite{mishchenko2006multiple,mishchenko2014electromagnetic}. The electromagnetic response after a long period of time or over a sufficient large spatial range can be computed by taking average of all possible configurations of particle distributions, or ensemble average \cite{mishchenko2006multiple,mishchenko2014electromagnetic}. The analytical definition of ensemble average is given in detail in Appendix \ref{en_avg_def}. The volume fraction of the identical spherical particles is $f_v$ and $n_0$ is the number density given by $n_0=3f_v/(4\pi a^3)$, where $a$ is the sphere radius. The random medium is also supposed to be statistically homogeneous and isotropic. We further restrict our model in the linear optics regime. Under these assumptions, the electromagnetic interaction of the incident light with the random medium is then described by the well-known FLEs, which can fully depict the many-particle multiple scattering process. The FLEs for $N$ particles read \cite{laxRMP1951,varadanJOSAA1985,mackowskiJOSAA1996,tsang2004scattering2}
\begin{equation}\label{fl_eq}
\mathbf{E}_{\text{exc}}^{(j)}(\mathbf{r})=\mathbf{E}_{\text{inc}}(\mathbf{r})+\sum_{i=1\atop i\neq j}^{N}\mathbf{E}_{\text{sca}}^{(i)}(\mathbf{r}),
\end{equation}
where $\mathbf{E}_{\text{inc}}(\mathbf{r})$ is the incident electric field, $\mathbf{E}_{\text{exc}}^{(j)}(\mathbf{r})$ is the electric component of the so-called exciting field impinging on the vicinity of particle $j$, and $\mathbf{E}_{\text{sca}}^{(i)}(\mathbf{r})$ is  electric component of partial scattered waves from particle $i$. We also show FLEs schematically in Fig.\ref{system_config} .

To solve this equation, it is convenient to expand the electric fields in VSWFs to utilize the spherical boundary condition of individual particles, following the way usually done for a single spherical particle. The expansion coefficients then naturally correspond to multipoles excited by the particles. This is a very mature scheme and has been widely used and extended by many authors \cite{lambPRB1980,tsangJAP1982,mackowskiJOSAA1996,tsang2004scattering,xuJQSRT2001}. Using this technique, the exciting field $\mathbf{E}_{\text{exc}}^{(j)}(\mathbf{r})$ is expressed as
\begin{equation}\label{exc_expansion}
\mathbf{E}_{\text{exc}}^{(j)}(\mathbf{r})=\sum_{mnp} c_{mnp}^{(j)}\mathbf{N}^{(1)}_{mnp}(\mathbf{r}-\mathbf{r}_j),
\end{equation}
where $\mathbf{N}^{(1)}_{mnp}(\mathbf{r}-\mathbf{r}_j)$ is the type-1 VSWF (or regular VSWF, defined in Appendix \ref{vswf_appendix}). The abbreviated summation $\sum_{mnp}$ generally stands for $\sum_{n=1}^{\infty}\sum_{m=-n}^{m=n}\sum_{p=1}^{2}$. Since here we only consider electric and magnetic dipole modes, it is possible for us to only take the first-order expansion into account, i.e., $n=1$. This is valid when the particles are not too densely packed that near-field coupling induces higher order multipoles. When $n=1$, the degree $m$ of VSWFs can only be $-1,0,1$. The subscript $p=1, 2$ denotes magnetic (TM) or electric (TE) modes respectively. Based on the expansion coefficients of exciting field, the scattering field from particle $i$ propagating to arbitrary position $\mathbf{r}$ can be obtained through its $T$-matrix elements $T_{np}$ as \cite{mackowskiJOSAA1996,tsang2004scattering}
\begin{equation}\label{sca_expansion}
\mathbf{E}_{\text{sca}}^{(i)}(\mathbf{r})=\sum_{n=1, mp} c_{mnp}^{(i)}T_{np}\mathbf{N}^{(3)}_{m1p}(\mathbf{r}-\mathbf{r}_i),
\end{equation}
where $\mathbf{N}^{(3)}_{mnp}(\mathbf{r}-\mathbf{r}_j)$ is the type-3 VSWF (or outgoing VSWF, defined in Appendix \ref{vswf_appendix}). For spherical particles $T$-matrix elements are the same as Mie coefficients, namely, $T_{12}=a_1$ and $T_{11}=b_1$ for dual-dipolar particles. Inserting Eqs.(\ref{exc_expansion}) and (\ref{sca_expansion}) into Eq.(\ref{fl_eq}), we obtain
\begin{equation}\label{fl_eq2}
\begin{split}
\sum_{mp} c_{m1p}^{(j)}\mathbf{N}^{(1)}_{m1p}(\mathbf{r}-\mathbf{r}_j)&=\mathbf{E}_{\text{inc}}(\mathbf{r})+\sum_{i=1\atop i\neq j}^{N}\sum_{mp} c_{m1p}^{(i)}T_{1p}\\&\cdot\mathbf{N}^{(3)}_{m1p}(\mathbf{r}-\mathbf{r}_i).
\end{split}
\end{equation}

To solve this equation, we need to translate the VSWFs centered at $\mathbf{r}_i$ to their counterparts centered at $\mathbf{r}_j$. Using translation addition theorem for VSWFs (see Appendix \ref{vswf_appendix}), Eq.(\ref{fl_eq2}) becomes
\begin{equation}\label{fl_eq3}
\begin{split}
\sum_{mp} c_{m1p}^{(j)}\mathbf{N}^{(1)}_{m1p}(\mathbf{r}-\mathbf{r}_j)&=\mathbf{E}_{\text{inc}}(\mathbf{r})+\sum_{i=1\atop i\neq j}^{N}\sum_{mp\mu q} c_{\mu 1q}^{(i)}T_{1q}\\&\cdot A_{mp\mu q}^{(3)}(\mathbf{r}_j-\mathbf{r}_i)\mathbf{N}^{(1)}_{m 1p}(\mathbf{r}-\mathbf{r}_j), 
\end{split}
\end{equation}
where $A_{mp\mu q}^{(3)}(\mathbf{r}_j-\mathbf{r}_i)$ can translate the outgoing VSWFs centered at $\mathbf{r}_i$ to regular VSWFs centered at $\mathbf{r}_j$. 
We further expand the incident waves into regular VSWFs centered at $\mathbf{r}_j$ with expansion coefficients $a_{mnp}^{(j)}$, use the orthogonal relation of VSWFs with different orders and degrees, and obtain the following equation:
\begin{equation}\label{fl_eq4}
c_{mp}^{(j)}=a_{mp}^{(j)}+\sum_{i=1\atop i\neq j}^{N}\sum_{\mu q}c_{\mu q}^{(i)}T_{1q}A_{mp\mu q}^{(3)}(\mathbf{r}_j-\mathbf{r}_i).
\end{equation}
Here $n=1$ in the subscript of $c_{mnp}^{(j)}$ and $a_{mnp}^{(j)}$ is omitted since only dipole excitations are of concern. To obtain the statistical averaged properties of electromagnetic wave propagation, we take ensemble average of Eq.(\ref{fl_eq4}) with respect to a fixed particle centered at $\mathbf{r}_j$ as
\begin{equation}\label{fl_avg_eq}
\langle c_{mp}^{(j)}\rangle_j=\langle a_{mp}^{(j)}\rangle_j+\Big\langle\sum_{i=1\atop i\neq j}^{N}\sum_{\mu q}c_{\mu q}^{(i)}T_{1q}A_{mp\mu q}^{(3)}(\mathbf{r}_j-\mathbf{r}_i)\Big\rangle_j,
\end{equation}
where $\langle \cdot \rangle_j$ denotes the ensemble average procedure with $\mathbf{r}_j$ fixed. Since for statistically homogeneous random media, ensemble average procedure restores the translational symmetry. The statistically averaged electromagnetic field in random media, or the coherent field, as proved by Lax \cite{laxPR1952}, is a plane wave. Here we only consider transverse electromagnetic wave propagation and assume the random medium only supports transverse coherent modes. We denote the effective propagation wave vector of the coherent wave by $\mathbf{K}$. Furthermore, it is assumed that the effective exciting field for particle $j$, which is equal to the total coherent field minus the field scattered by the investigated scatterer $j$, is also planewave-like possessing the same wave vector, but with a different amplitude \cite{laxPR1952}.
\begin{equation}\label{aprox1}
\langle c_{mp}^{(j)}\rangle_j\approx C_{mp}\exp({i\mathbf{K}\cdot\mathbf{r}_j}),
\end{equation}
where $C_{mp}$ is the expansion coefficient of effective exciting wave amplitude at the origin, which has the same physical significance of Lax's effective field factor \cite{laxRMP1951,laxPR1952}. The expansion coefficient for different particles only differ by a plane-wave type phase shift, and $C_{mp}$ only depends on the overall property of the random media. According to the principle of modal analysis (MA), for passive media (the present case), the solution of propagation constant  $K=|\mathbf{K}|$ can be found in the upper complex plane to meet the poles of Green's function for coherent wave propagation \cite{campionePNFA2013,sheng2006introduction}. In other words, the effective propagation constant $K$ corresponds to the most probable mode with the maximal response and minimal extinction in the random media, namely, the mode with the largest spectral function \cite{sheng2006introduction}. To solve Eq.(\ref{fl_avg_eq}), the expression for $\langle c_{\mu q}^{(i)}T_{1q}A_{\mu qmp}^{(3)}(\mathbf{r}_i-\mathbf{r}_j)\rangle_j$ should be given first. Herein the QCA suggested by Lax \cite{laxPR1952} is introduced, which expresses high-order correlations among three or more particles using two-particle statistics and amounts to
\begin{equation}\label{aprox2}
\langle c_{mp}^{(i)}\rangle_{ij}\approx\langle c_{mp}^{(i)}\rangle_{i}\approx C_{mp}\exp({i\mathbf{K}\cdot\mathbf{r}_i}),
\end{equation}
where $\langle \cdot \rangle_{ij}$ denotes the ensemble average procedure with $\mathbf{r}_j$ and $\mathbf{r}_i$ fixed simultaneously. This equation suggests that the fluctuation of the effective exciting field impinging on particle \textit{i} due to a deviation of particle \textit{j} from its average position can be neglected. This is strictly valid for a periodic or crystalline medium, but to some extent is viable for a densely packed medium possessing partial order, as suggested by Lax \cite{laxPR1952}. Inserting Eqs.(\ref{aprox1}-\ref{aprox2}) into Eq.(\ref{fl_avg_eq}) and using the definition of PDF in Appendix \ref{en_avg_def}, we obtain
%\begin{equation}
%C_{mp}=\langle a_{mp}^{j}\rangle+\Big\langle\sum_{i=1\atop i\neq j}^{N}\sum_{\mu q}C_{\mu q}T_{1q}A_{\mu qmp}^{(3)}(\mathbf{r}_i,\mathbf{r}_j)\exp({i\mathbf{K}\cdot\mathbf{r}_i-i\mathbf{K}\cdot\mathbf{r}_j})\Big\rangle
%\end{equation}
\begin{equation}\label{fl_avg_eq2}
\begin{split}
C_{mp}&=\langle a_{mp}^{(j)}\rangle_j\exp({-i\mathbf{K}\cdot\mathbf{r}_j})+\sum_{\mu q}C_{\mu q}T_{1q}\\&\cdot\Big\langle\sum_{i=1\atop i\neq j}^{N}A_{mp\mu q}^{(3)}(\mathbf{r}_j-\mathbf{r}_i)\exp[{i\mathbf{K}(\mathbf{r}_i-\mathbf{r}_j)}]\Big\rangle_j\\
&=\langle a_{mp}^{(j)}\rangle_j\exp({-i\mathbf{K}\cdot\mathbf{r}_j})+n_0\sum_{\mu q}C_{\mu q}T_{1q}\\&\cdot\int d\mathbf{r}A_{mp\mu q}^{(3)}(-\mathbf{r})\exp{(i\mathbf{K}\cdot\mathbf{r})}g_2(\mathbf{r}),
\end{split}
\end{equation}
where $\mathbf{r}=\mathbf{r}_i-\mathbf{r}_j$. Here we use the fact that for the present statistically homogeneous medium with translational invariance, $g_2(\mathbf{r}_i,\mathbf{r}_j)$ only depends on $\mathbf{r}=\mathbf{r}_i-\mathbf{r}_j$. Furthermore, for the considered isotropic medium, $g_2(\mathbf{r})=g_2(r)$, where $r=|\mathbf{r}|$.  

For convenience, we express $C_{mp}$ into a column vector $\tilde{\mathbf{C}}$, where the elements $\tilde{C}_\alpha$ are numbered as $\alpha=1,2,3,4,5,6$, denoting different combinations of $mp$, i.e., $\alpha=1,2,3,4,5,6 \rightarrow (m,p)= (-1,1),(0,1),(1,1),(-1,2),(0,2),(1,2)$. Similarly, we use $\tilde{A}_{\beta\alpha}(\mathbf{K})$ to denote the integral containing translation coefficient, which is a function of the unknown effective propagation constant $\mathbf{K}$, as 
\begin{equation}\label{translation_integral}
\tilde{A}_{mp\mu q}(\mathbf{K})=\int d\mathbf{r}A_{mp\mu q}^{(3)}(-\mathbf{r})\exp{(i\mathbf{K}\cdot\mathbf{r})}g_2(\mathbf{r})
\end{equation}
for $\alpha=(\mu,q)$ and $\beta=(m,p)$. And the diagonal elements of the $T$-matrix is given by $\tilde{T}_{11}=\tilde{T}_{22}=\tilde{T}_{33}=T_{11}=b_1$ and $\tilde{T}_{44}=\tilde{T}_{55}=\tilde{T}_{66}=T_{12}=a_1$, in which other elements are all zeros. The incident wave amplitudes are expressed in a column vector $\tilde{\mathbf{a}}$ with $\tilde{a}_{\alpha}=\langle a_{mp}^{j}\rangle_j\exp({-i\mathbf{K}\cdot\mathbf{r}_j})$. Therefore we have the final solution as
\begin{equation}\label{matrix_eq1}
\tilde{\mathbf{C}}=\tilde{\mathbf{a}}+n_0\tilde{\mathbf{A}}(\mathbf{K})\tilde{\mathbf{T}}\tilde{\mathbf{C}}.
\end{equation}
Thus $\mathbf{C}$ is solved by
\begin{equation}\label{matrix_eq2}
\mathbf{\tilde{C}}=[\mathbf{I}-n_0\tilde{\mathbf{A}}(\mathbf{K})\tilde{\mathbf{T}}]^{-1}\tilde{\mathbf{a}}.
\end{equation}
According to the definition of the most probable propagating mode, for arbitrary incident wave $\tilde{\mathbf{a}}$, $K$ should meet the pole of $[\mathbf{I}-n_0\tilde{\mathbf{A}}(\mathbf{K})\tilde{\mathbf{T}}]^{-1}$. This amounts to finding the $K$ leads to the determinant $|\mathbf{I}-n_0\tilde{\mathbf{A}}(\mathbf{K})\tilde{\mathbf{T}}|$ to be zero in the upper complex plane for passive media under consideration \cite{sheng2006introduction,campionePNFA2013}. In the next we will evaluate the matrix elements of $\tilde{\mathbf{A}}$ and solve the effective propagation constant $K$.

In spherical coordinates, the translation coefficients are given in terms of the position vector $\mathbf{r}=(r,\theta,\phi)$ as \cite{tsang2004scattering,mackowskiJOSAA1996,bringi1982coherent}
\begin{equation}\label{translation_coef1}
\begin{split}
&A_{m1\mu 1}^{(3)}(\mathbf{r})=A_{m2\mu 2}^{(3)}(\mathbf{r})=\sqrt{\frac{(1-\mu)!(1+m)!}{(1+\mu)!(1-m)!}}(-1)^{m}\\&\cdot\sum_{n}a(\mu,1|-m,1|n)a(1,1,n)h_n(kr)Y_{n}^{\mu-m}(\theta,\phi),
\end{split}
\end{equation}
\begin{equation}\label{translation_coef2}
\begin{split}
&A_{m1\mu 2}^{(3)}(\mathbf{r})=A_{m2\mu 1}^{(3)}(\mathbf{r})=\sqrt{\frac{(1-\mu)!(1+m)!}{(1+\mu)!(1-m)!}}(-1)^{m+1}\\&\sum_{n}a(\mu,1|-m,1|n,n-1)b(1,1,n)h_n(kr)Y_{n}^{\mu-m}(\theta,\phi),
\end{split}
\end{equation}
where $Y_n^{m}(\theta,\phi)$ are spherical harmonics, and coefficients that contain $m, \mu,n$ including $a(m,1|-\mu,1|n)$, $a(\mu,1|-m,1|n)$, $a(1,1,n)$, $b(1,1,n)$ are related to Wigner 3-$j$ symbols and listed in Appendix \ref{vswf_appendix}. For dual dipolar excitations, $n$ can be only chosen to be $0,1,2$ to make these coefficients nonzero \cite{tsang2004scattering}.
Incorporating Eqs.(\ref{translation_coef1}) and (\ref{translation_coef2}) into Eq.(\ref{translation_integral}) and invoking the well-known plane wave expansion \cite{abramowitz1964handbook}
\begin{equation}
\begin{split}
\exp{(i\mathbf{K}\cdot\mathbf{r})}=\sum_{l,m}(2l+1)i^lj_l(Kr)Y_l^{m*}(\theta,\phi)Y_l^m(\theta_K,\phi_K),
\end{split}
\end{equation}
we can obtain a typical integral in Eq.(\ref{translation_integral}) as follows:
\begin{equation}
\begin{split}
I_{n}^{m,\mu}(\mathbf{K})&=\int d\mathbf{r}(-1)^nh_n(kr)Y_{n}^{\mu-m}(\theta,\phi)\sum_{n',m'}(2n'+1)i^{n'}\\&\cdot j_{n'}(Kr)Y_{n'}^{m'*}(\theta,\phi)Y_{n'}^{m'}(\theta_K,\phi_K)g_2(r),
\end{split}
\end{equation}
which can be evaluated numerically for a known $g_2(r)$. In the present paper, all the integrals, if not specified, are performed over the entire real (for position vector) or reciprocal (for reciprocal vector) spaces. Using the orthogonal relation of spherical harmonics \cite{abramowitz1964handbook},
\begin{equation}
\int_{4\pi} Y_{n}^{m}(\theta,\phi)Y_{n'}^{m'*}(\theta,\phi)d\varOmega=\frac{4\pi}{2n+1}\delta_{nn'}\delta_{mm'},
\end{equation}
where $\varOmega$ is the solid angle, above integral becomes
\begin{equation}
\begin{split}
I_{n}^{m,\mu}(\mathbf{K})=\int_0^\infty 4\pi(-i)^nh_n(kr)j_{n}(Kr)Y_{n}^{\mu-m}(\theta_K,\phi_K)g_2(r)r^2dr.
\end{split}
\end{equation}
As a consequence, the matrix elements of $\tilde{\mathbf{A}}(\mathbf{K})$ are obtained as
\begin{equation}
\begin{split}
&\tilde{A}_{m1\mu 1}(\mathbf{K})=\tilde{A}_{m2\mu 2}(\mathbf{K})=\sqrt{\frac{(1-\mu)!(1+m)!}{(1+\mu)!(1-m)!}}(-1)^{m}\\&\cdot\sum_{n}a(\mu,1|-m,1|n)a(1,1,n)I_{n}^{m,\mu}(\mathbf{K}),
\end{split}
\end{equation}
\begin{equation}
\begin{split}
&\tilde{A}_{m2\mu 1}(\mathbf{K})=\tilde{A}_{m1\mu 2}(\mathbf{K})=\sqrt{\frac{(1-\mu)!(1+m)!}{(1+\mu)!(1-m)!}}(-1)^{m+1}\\&\cdot\sum_{n}a(\mu,1|-m,1|n,n-1)b(1,1,n)I_{n}^{m,\mu}(\mathbf{K}).
\end{split}
\end{equation}

Without loss of generality, we assume that the propagation direction of incident and coherent waves is the $z$-axis as shown in Fig.\ref{system_config}, which leads to \cite{abramowitz1964handbook}
\begin{equation}
Y_{n}^{\mu-m}(\theta_K=0,\phi_K)=\delta_{m-\mu,0},
\end{equation}
which demands $m=\mu$. In this circumstance, Therefore, $\tilde{\mathbf{A}}$ only depends on $K$, and the nonzero elements of $\tilde{\mathbf{A}}(K)$ are only $\tilde{A}_{11}(K)=\tilde{A}_{44}(K)$, $\tilde{A}_{22}(K)=\tilde{A}_{55}(K)$, $\tilde{A}_{33}(K)=\tilde{A}_{66}(K)$, $\tilde{A}_{14}(K)=\tilde{A}_{41}(K)$, $\tilde{A}_{25}(K)=\tilde{A}_{52}(K)$, $\tilde{A}_{36}(K)=\tilde{A}_{63}(K)$. The condition $|\mathbf{I}-n_0\tilde{\mathbf{A}(K)}\tilde{\mathbf{T}}|=0$ is also equivalent to $(\mathbf{I}-n_0\tilde{\mathbf{A}(K)}\tilde{\mathbf{T}})\tilde{\mathbf{C}}=\mathbf{0}$. After some manipulations, we obtain
\begin{equation}
\begin{split}
\left(\begin{matrix}
(1-n_0\tilde{A}_{11}(K)b_1)\tilde{C}_1-n_0\tilde{A}_{41}(K)a_1\tilde{C}_4\\
(1-n_0\tilde{A}_{22}(K)b_1)\tilde{C}_2-n_0\tilde{A}_{52}(K)a_1\tilde{C}_5\\
(1-n_0\tilde{A}_{33}(K)b_1)\tilde{C}_3-n_0\tilde{A}_{63}(K)a_1\tilde{C}_6
\end{matrix}\right)=\mathbf{0}
\end{split}
\end{equation}
and
\begin{equation}
\begin{split}
\left(\begin{matrix}
(1-n_0\tilde{A}_{11}(K)a_1)\tilde{C}_4-n_0\tilde{A}_{41}(K)b_1\tilde{C}_1\\
(1-n_0\tilde{A}_{22}(K)a_1)\tilde{C}_5-n_0\tilde{A}_{52}(K)b_1\tilde{C}_2\\
(1-n_0\tilde{A}_{33}(K)a_1)\tilde{C}_6-n_0\tilde{A}_{63}(K)b_1\tilde{C}_3
\end{matrix}\right)=\mathbf{0}.
\end{split}
\end{equation}
%Therefore we have
%\begin{equation}
%\frac{1-n_0B_{11}b_1}{n_0B_{41}a_1}=\frac{n_0B_{41}b_1}{1-n_0B_{11}a_1}
%\end{equation}
%\begin{equation}
%\frac{1-n_0B_{22}b_1}{n_0B_{52}a_1}=\frac{n_0B_{52}b_1}{1-n_0B_{22}a_1}
%\end{equation}
%\begin{equation}
%\frac{1-n_0B_{33}b_1}{n_0B_{63}a_1}=\frac{n_0B_{63}b_1}{1-n_0B_{33}a_1}
%\end{equation}
These equations are still a little bit complicated to solve. However, we can simplify the solution by only considering the plane wave illumination. Without loss of generality, the incident plane wave is assumed to be linearly polarized over the $y$-axis and propagate along $z$-axis with unity amplitude, namely, $\mathbf{E}_{\text{inc}}=\hat{\mathbf{y}}\exp({ikz})$, which therefore can be expanded in regular VSWFs centered at $\mathbf{r}_j$ as \cite{mackowskiJOSAA1996,tsang2004scattering2}
\begin{equation}
\mathbf{E}_{\text{inc}}(\mathbf{r})=\exp({i\mathbf{k}\cdot \mathbf{r}_j})\sum_{np}\sum_{m=\pm1} a_{mnp}^{0}\mathbf{N}^{(1)}_{mnp}(\mathbf{r}-\mathbf{r}_j),
\end{equation}
where 
\begin{equation}\label{pw_coef1}
a_{1n1}^{0}=-a_{-1n1}^{0}=\frac{1}{2}[4\pi(2n+1)]^{1/2},
\end{equation}
\begin{equation}\label{pw_coef2}
a_{1n2}^{0}=a_{-1n2}^{0}=\frac{1}{2}[4\pi(2n+1)]^{1/2},
\end{equation}
where the coefficients corresponding to $m=0$ are all equal to zero. Hence $\langle a_{mnp}^{(j)}\rangle=a_{mnp}^{(j)}=\exp({i\mathbf{k}\cdot \mathbf{r}_j})a_{mnp}^{0}$, where $a_{mnp}^{0}$ is the expansion coefficient of incident wave at the origin. Regarding Eqs.(\ref{pw_coef1}) and (\ref{pw_coef2}), for the linearly polarized coherent wave, we also have $\tilde{C}_{1}=-\tilde{C}_{3}$, $\tilde{C}_{2}=\tilde{C}_{5}=0$ and $\tilde{C}_{4}=\tilde{C}_{6}$.
%\begin{equation}
%(1-n_0B_{11}b_1)C_1-n_0B_{41}a_1C_4+(1-n_0B_{33}b_1)C_3-n_0B_{63}a_1C_6=0
%\end{equation}
%\begin{equation}
%(1-n_0B_{11}a_1)C_4-n_0B_{41}b_1C_1=(1-n_0B_{33}a_1)C_6-n_0B_{63}b_1C_3
%\end{equation}
%\begin{equation}
%(1-n_0B_{22}b_1)C_2-n_0B_{52}a_1C_5=0
%\end{equation}
%\begin{equation}
%(1-n_0B_{22}a_1)C_5-n_0B_{52}b_1C_2=0
%\end{equation}
%This gives us
%\begin{equation}
%\begin{split}
%\left(\begin{matrix}
%-(1-n_0\tilde{A}_{11}b_1)\tilde{C}_3-n_0\tilde{A}_{41}a_1\tilde{C}_6\\
%(1-n_0\tilde{A}_{33}b_1)\tilde{C}_3-n_0\tilde{A}_{63}a_1\tilde{C}_6
%\end{matrix}\right)=\mathbf{0}
%\end{split}
%\end{equation}
%And
%\begin{equation}
%\begin{split}
%\left(\begin{matrix}
%(1-n_0\tilde{A}_{11}a_1)\tilde{C}_6+n_0\tilde{A}_{41}b_1\tilde{C}_3\\
%(1-n_0\tilde{A}_{33}a_1)\tilde{C}_6-n_0\tilde{A}_{63}b_1\tilde{C}_3
%\end{matrix}\right)=\mathbf{0}
%\end{split}
%\end{equation}
Moreover, since $a(1,1|-1,1|n)=a(-1,1|1,1|n)$ (non zero for $n=0,2$), we have $\tilde{A}_{11}(K)=\tilde{A}_{33}(K)$, and $a(1,1|-1,1|n,n-1)=-a(-1,1|1,1|n,n-1)$ (non zero only for $n=1$) leads to $\tilde{A}_{41}(K)=-\tilde{A}_{63}(K)$ (refer to Appendix \ref{vswf_appendix}), we are finally able to obtain the following equations containing only two unknowns $\tilde{C}_3$ and $\tilde{C}_6$ as
\begin{equation}\label{c_eq1}
\begin{split}
(1-n_0\tilde{A}_{33}(K)b_1)\tilde{C}_3-n_0\tilde{A}_{63}(K)a_1\tilde{C}_6=0,
\end{split}
\end{equation}
\begin{equation}\label{c_eq2}
\begin{split}
(1-n_0\tilde{A}_{33}(K)a_1)\tilde{C}_6-n_0\tilde{A}_{63}(K)b_1\tilde{C}_3=0.
\end{split}
\end{equation}
Therefore we get the final dispersion relation as
\begin{equation}\label{dispersion_relate_eq}
(1-n_0\tilde{A}_{33}(K)b_1)(1-n_0\tilde{A}_{33}(K)a_1)-n_0^2\tilde{A}_{63}^2a_1b_1=0,
\end{equation}
where the effective propagation constant $K$ can be solved in the upper complex plane.

Since it is not possible to solve the effective exciting field amplitudes $C_{mp}$ with only Eqs.(\ref{c_eq1}) and (\ref{c_eq2}), in the next, we will derive an extra equation for this purpose.  Note $\tilde{C}_3$ and $\tilde{C}_6$ are not necessarily equal following from a plane wave, because the electric and magnetic responses of the particles might not be the same. To solve them, we consider the relationship between the effective propagation constant and the transmission coefficient of the coherent field. For the coherent field, it propagates in the random medium in a ballistic manner, the same as the case in a homogeneous medium. The transmission coefficient, for the coherent field normally illuminated onto a random medium slab with an effective propagation constant $\mathbf{K}=K\hat{z}$ and thickness $w$, is calculated as \cite{laxPR1952,bornandwolf,mackowskiJQSRT2013}
\begin{equation}\label{t_slab}
t=1+\frac{iw(K^2-k^2)}{2k}.
\end{equation}
On the other hand, the transmission coefficient for the random medium slab can be calculated from the ensemble averaged, far-field forward scattering amplitude $f(\mathbf{K},\mathbf{K})$ \cite{laxPR1952}, where the first $\mathbf{K}$ indicates the wave vector of the incident field and the second stands for that of the scattered field. To this end, we first consider the total scattered field that is given by
\begin{equation}
\begin{split}
\langle\mathbf{E}_\text{s}(\mathbf{r})\rangle=\langle\sum_{j=1}^{N}\mathbf{E}_\text{s}^{(j)}(\mathbf{r})\rangle&=n_0\int d\mathbf{r}_j\sum_{mp}\langle c_{m1p}^{(j)}\rangle T_{1p}\\&\cdot\mathbf{N}^{(3)}_{m1p}(\mathbf{r}-\mathbf{r}_j).
\end{split}
\end{equation}
Taking far-field approximation ($r\rightarrow\infty$) for outgoing VSWFs (see Appendix \ref{far-field_appendix}), after some manipulations, we have
\begin{equation}
\langle\mathbf{E}_\text{s}(\mathbf{r})\rangle=\frac{n_0\exp{(ikr)}}{r}\frac{3i}{2k}(a_1C_{12}+b_1C_{11}).
\end{equation}
Therefore the forward scattering amplitude is obtained as
\begin{equation}
f(\mathbf{K},\mathbf{K})=\frac{3in_0}{2k}(a_1C_{12}+b_1C_{11}).
\end{equation}
And the transmission coefficient is related to forward scattering amplitude $f(\mathbf{K},\mathbf{K})$ through  
\begin{equation}\label{t_random}
t=1+\frac{2\pi iwf(\mathbf{K},\mathbf{K})}{k}.
\end{equation}
A comparison between Eqs.(\ref{t_slab}) and (\ref{t_random}) leads to the following relation
\begin{equation}\label{extionction_theorem}
K^2-k^2=\frac{6\pi in_0}{k}(a_1C_{12}+b_1C_{11}).
\end{equation}
Combining Eqs.(\ref{c_eq1}),(\ref{dispersion_relate_eq}) and (\ref{extionction_theorem}), the effective exciting field amplitudes $C_{12}$ and $C_{11}$ can be solved. 

By this stage, we have established the theory for electromagnetic field propagation in a random medium consisting of dual-dipolar particles, where the formulas of the dispersion relation and effective exciting field amplitudes are derived under Lax's QCA. This theory provides an analytical tool for examining the interplay between the dipolar modes of a single scatterer and the structural correlations.

\subsection{Scattering phase function}
After calculating the effective propagation constant for the coherent wave, to derive the scattering phase function defined for incoherent waves \cite{maAO1988}, it is necessary to compute the scattering intensity in this random medium. We first consider the coherent field that is the ensemble averaged total field:
\begin{equation}
\mathbf{E}_{\text{coh}}(\mathbf{r})=\langle\mathbf{E}(\mathbf{r})\rangle=\langle\mathbf{E}_{\text{inc}}(\mathbf{r})+\mathbf{E}_\text{s}(\mathbf{r})\rangle,
\end{equation}
where $\mathbf{E}(\mathbf{r})=\mathbf{E}_{\text{inc}}(\mathbf{r})+\mathbf{E}_\text{s}(\mathbf{r})$ is the total field. The coherent intensity is then defined as $I_{\text{coh}}(\mathbf{r})=\mathbf{E}_{\text{coh}}(\mathbf{r})\mathbf{E}_{\text{coh}}^*(\mathbf{r})$, where the superscript $*$ denotes the complex conjugate, and the ensemble averaged total intensity is given by
\begin{equation}
I(\mathbf{r})=\langle\mathbf{E}(\mathbf{r})\mathbf{E}^*(\mathbf{r})\rangle=\langle[\mathbf{E}_{\text{inc}}(\mathbf{r})+\mathbf{E}_\text{s}(\mathbf{r})]\cdot[\mathbf{E}_{\text{inc}}^*(\mathbf{r})+\mathbf{E}_\text{s}^*(\mathbf{r})]\rangle.
\end{equation}
Therefore the incoherent intensity, defined as the difference between total intensity and coherent intensity, is calculated through
\begin{equation}
I_{\text{ich}}(\mathbf{r})=I(\mathbf{r})-I_\text{coh}(\mathbf{r})=\langle\mathbf{E}_\text{s}(\mathbf{r})\mathbf{E}_\text{s}^*(\mathbf{r})\rangle-\langle\mathbf{E}_\text{s}(\mathbf{r})\rangle\langle\mathbf{E}_\text{s}^*(\mathbf{r})\rangle,
\end{equation}
which describes the light intensity generated by random fluctuations of the medium, also known as the diffuse intensity. \cite{mackowskiJQSRT2013} To proceed, we write down the ensemble averaged intensity of scattered wave in above equation as 
\begin{equation}\label{total_sca_eq}
\begin{split}
&\langle\mathbf{E}_\text{s}(\mathbf{r})\mathbf{E}_\text{s}^*(\mathbf{r})\rangle=n_0\sum_{mpm'p'}\int d\mathbf{r}_j\mathbf{N}^{(3)}_{m1p}(\mathbf{r}-\mathbf{r}_j)\mathbf{N}^{(3)*}_{m1p}(\mathbf{r}-\mathbf{r}_j)\\&\cdot T_{1p}T_{1p'}^*\langle c_{mp}^{(j)}c_{m'p'}^{(j)*}\rangle_j+n_0^2\sum_{mpm'p'}\iint d\mathbf{r}_j d\mathbf{r}_ig_2(\mathbf{r}_j-\mathbf{r}_i)\\&\cdot T_{1p}T_{1p'}^*\langle c_{mp}^{(j)}c_{m'p'}^{(i)*}\rangle_{ij}\mathbf{N}^{(3)}_{m1p}(\mathbf{r}-\mathbf{r}_j)\mathbf{N}^{(3)*}_{m1p}(\mathbf{r}-\mathbf{r}_i).
\end{split}
\end{equation}
In the meanwhile, the coherent scattered intensity is given by
\begin{equation}\label{coh_sca_eq}
\begin{split}
&\langle\mathbf{E}_\text{s}(\mathbf{r})\rangle\langle\mathbf{E}_\text{s}^*(\mathbf{r})\rangle=n_0^2\sum_{mpm'p'}\iint d\mathbf{r}_j d\mathbf{r}_iT_{1p}T_{1p'}^*\\&\cdot \langle c_{mp}^{(j)}\rangle_{j}\langle c_{m'p'}^{(i)*}\rangle_{i}\mathbf{N}^{(3)}_{m1p}(\mathbf{r}-\mathbf{r}_j)\mathbf{N}^{(3)*}_{m1p}(\mathbf{r}-\mathbf{r}_i).
\end{split}
\end{equation}
The unknowns in Eq.(\ref{total_sca_eq}) are $\langle c_{mp}^{(j)}c_{m'p'}^{(j)*}\rangle_j$ and $\langle c_{mp}^{(j)}c_{m'p'}^{(i)*}\rangle_{ij}$, in which the former is the intensity of exciting field with respect to a particular particle, and the latter stands for the correlation between the exciting fields impinging on different particles. The incoherent  intensity is therefore given by
\begin{equation}\label{incoh_sca_eq2}
\begin{split}
&I_{\text{ich}}(\mathbf{r})=n_0\sum_{mpm'p'}\iint d\mathbf{r}_j d\mathbf{r}_i[n_0g_2(\mathbf{r}_j-\mathbf{r}_i)+\delta(\mathbf{r}_j-\mathbf{r}_i)] \\&\cdot T_{1p}T_{1p'}^*\mathbf{N}^{(3)}_{m1p}(\mathbf{r}-\mathbf{r}_j)\mathbf{N}^{(3)*}_{m1p}(\mathbf{r}-\mathbf{r}_i) \langle c_{mp}^{(j)}c_{m'p'}^{(i)*}\rangle_{ij}\\&-n_0^2\sum_{mpm'p'}\iint d\mathbf{r}_jd\mathbf{r}_iT_{1p}T_{1p'}^*\langle c_{mp}^{(j)}\rangle_{j}\langle c_{m'p'}^{(i)*}\rangle_{i} \\&\cdot \mathbf{N}^{(3)}_{m1p}(\mathbf{r}-\mathbf{r}_j)\mathbf{N}^{(3)*}_{m1p}(\mathbf{r}-\mathbf{r}_i).
\end{split}
\end{equation}
where Dirac delta function $\delta(\mathbf{r}_j-\mathbf{r}_i)$ takes the case when $\mathbf{r}_j$ and $\mathbf{r}_i$ stand for the same particle. In the spirit of QCA we make an assumption for the correlation $\langle c_{mp}^{(j)}c_{m'p'}^{(i)*}\rangle_{ij}$, which is similar to Lax's method using the so-called effective field factor \cite{laxRMP1951}:
\begin{equation}\label{qca_intensity_assumption}
\langle c_{mp}^{(j)}c_{m'p'}^{(i)*}\rangle_{ij}\approx C_{mp}C_{m'p'}^*\langle \mathbf{E}(\mathbf{r}_j)\mathbf{E}^*(\mathbf{r}_i)\rangle.
\end{equation}
Inserting Eqs. (\ref{aprox2}) and (\ref{qca_intensity_assumption}) into Eq.(\ref{incoh_sca_eq2}), we obtain
\begin{equation}\label{incoh_sca_eq3}
\begin{split}
&I_{\text{ich}}(\mathbf{r})=n_0\sum_{mpm'p'}\iint d\mathbf{r}_j d\mathbf{r}_i[n_0h_2(\mathbf{r}_j-\mathbf{r}_i)+\delta(\mathbf{r}_j-\mathbf{r}_i)]\\&\cdot T_{1p}T_{1p'}^*C_{mp}C_{m'p'}^*\mathbf{N}^{(3)}_{m1p}(\mathbf{r}-\mathbf{r}_j)\mathbf{N}^{(3)*}_{m1p}(\mathbf{r}-\mathbf{r}_i)\langle \mathbf{E}(\mathbf{r}_j)\mathbf{E}^*(\mathbf{r}_i)\rangle\\&+n_0^2\sum_{mpm'p'}\iint d\mathbf{r}_j d\mathbf{r}_iT_{1p}T_{1p'}^* C_{mp}C_{m'p'}^*\mathbf{N}^{(3)}_{m1p}(\mathbf{r}-\mathbf{r}_j)\\&\cdot\mathbf{N}^{(3)*}_{m1p}(\mathbf{r}-\mathbf{r}_i)\Big(\langle \mathbf{E}(\mathbf{r}_j)\mathbf{E}^*(\mathbf{r}_i)\rangle-\langle \mathbf{E}(\mathbf{r}_j)\rangle\langle\mathbf{E}^*(\mathbf{r}_i)\rangle\Big).
\end{split}
\end{equation}
where $h_2(\mathbf{r})=g_2(\mathbf{r})-1$ is the pair correlation function. The first term in the right hand side (RHS) of Eq.(\ref{incoh_sca_eq3}) gives the incoherent intensity produced by the total intensity (specifically, the total field correlation $\langle \mathbf{E}(\mathbf{r}_j)\mathbf{E}^*(\mathbf{r}_i)\rangle$, which can be understood as a generalization of intensity \cite{sheng2006introduction,vynckPRA2016}), while the second term denotes the incoherent intensity generated only by incoherent intensity. Since the incoherent intensity, originated from random fluctuations of total intensity, is usually much smaller than total intensity \cite{sheng2006introduction,mackowskiJQSRT2013}, it is then possible for us to neglect this term in the RHS of Eq.(\ref{incoh_sca_eq3}). This assumption gives rise to 
\begin{equation}\label{incoh_sca_eq4}
\begin{split}
&I_{\text{ich}}(\mathbf{r})\approx n_0\sum_{mpm'p'}\iint d\mathbf{r}_j d\mathbf{r}_i[n_0h_2(\mathbf{r}_j-\mathbf{r}_i)+\delta(\mathbf{r}_j-\mathbf{r}_i)]\\&\cdot T_{1p}T_{1p'}^*C_{mp}C_{m'p'}^*\mathbf{N}^{(3)}_{m1p}(\mathbf{r}-\mathbf{r}_j)\mathbf{N}^{(3)*}_{m1p}(\mathbf{r}-\mathbf{r}_i)\langle \mathbf{E}(\mathbf{r}_j)\mathbf{E}^*(\mathbf{r}_i)\rangle.
\end{split}
\end{equation}
If we only consider first order scattering, i.e., $\langle \mathbf{E}(\mathbf{r}_j)\mathbf{E}^*(\mathbf{r}_i)\rangle\approx\langle \mathbf{E}(\mathbf{r}_j)\rangle\langle\mathbf{E}^*(\mathbf{r}_i)\rangle$, above equation becomes the well-known distorted Born approximation (DBA) for calculating radiative transfer in the remote sensing community \cite{twerskyJASA1957,tsangRS2000,maAO1988,tsang2004scattering}. However, our present equation reproduces all orders of multiple scattering if the total intensity $I(\mathbf{r})$ is repeatedly iterated. Eq.(\ref{incoh_sca_eq4}) is actually in the form of Bethe-Salpeter equation for wave propagation in random media \cite{barabanenkovJEWA1995,lagendijk1996resonant,VanRossum1998,tsang2004scattering,leseurOptica2016,sheng2006introduction,Cherroret2016}. The irreducible vertex governing the multiple scattering process is given by $\Gamma_{mpm'p'}=T_{1p}T_{1p'}^*C_{mp}C_{m'p'}^*[n_0h_2(\mathbf{r}_j-\mathbf{r}_i)+\delta(\mathbf{r}_j-\mathbf{r}_i)]$, which corresponds to a modfied ladder approximation accounting for particle correlations \cite{tsangJEWA1987,barabanenkovPLA1992}.

To derive the scattering phase function, the integral in Eq.(\ref{incoh_sca_eq4}) is needed to be carried out. For doing so, we express the field correlation function in its Fourier transform component in reciprocal (or momentum) space
\begin{equation}\label{ftrans_eq}
\begin{split}
\langle \mathbf{E}(\mathbf{r}_j)\mathbf{E}^*(\mathbf{r}_i)\rangle&=\int d\mathbf{p}_1d\mathbf{p}_2\langle \mathbf{E}(\mathbf{p}_1)\mathbf{E}^*(\mathbf{p}_2)\rangle\\&\cdot\exp{(i\mathbf{p}_1\cdot\mathbf{r}_j-i\mathbf{p}_2\cdot\mathbf{r}_i)}.
\end{split}
\end{equation}
%Since the random medium is statistically homogeneous and obeys translation symmetry, the Fourier component of field correlation function only depends on a single momentum $\mathbf{p}$.
where $\mathbf{p}_1$ and $\mathbf{p}_2$ are both reciprocal vectors, corresponding to $\mathbf{r}_j$ and $\mathbf{r}_i$ respectively.  Substituting Eq.(\ref{ftrans_eq}) into Eq.(\ref{incoh_sca_eq4}), we have
\begin{equation}
\begin{split}
&I_{\text{ich}}(\mathbf{r})=n_0\sum_{mpm'p'}\iiiint d\mathbf{r}_j d\mathbf{r}_id\mathbf{p}_1d\mathbf{p}_2[n_0h_2(\mathbf{r}_j-\mathbf{r}_i)\\&+\delta(\mathbf{r}_j-\mathbf{r}_i)] T_{1p}T_{1p'}^*C_{mp}C_{m'p'}^*\iint \mathbf{N}^{(3)}_{m1p}(\mathbf{p}_3)\mathbf{N}^{(3)*}_{m1p}(\mathbf{p}_4)\\&\cdot\langle \mathbf{E}(\mathbf{p}_1)\mathbf{E}^*(\mathbf{p}_2)\rangle\exp{(i\mathbf{p}_1\cdot\mathbf{r}_j-i\mathbf{p}_2\cdot\mathbf{r}_i)}\exp(i\mathbf{p}_3\cdot(\mathbf{r}-\mathbf{r}_j))\\&\cdot\exp(-i\mathbf{p}_4\cdot(\mathbf{r}-\mathbf{r}_i))d\mathbf{p}_3d\mathbf{p}_4.
\end{split}
\end{equation}
Here we also use the Fourier transform of VSWFs, $\mathbf{N}^{(3)}_{m1p}(\mathbf{p}_3)$ and $\mathbf{N}^{(3)*}_{m1p}(\mathbf{p}_4)$, with a similar definition to Eq.(\ref{ftrans_eq}), where $\mathbf{p}_3$ and $\mathbf{p}_4$ are also reciprocal vectors, corresponding to $\mathbf{r}-\mathbf{r}_j$ and $\mathbf{r}-\mathbf{r}_i$ respectively. To carry out above integral, we further change the variables as $\mathbf{R}=(\mathbf{r}_j+\mathbf{r}_i)/2$, $\mathbf{s}=\mathbf{r}_j-\mathbf{r}_i$, $\mathbf{p}=(\mathbf{p}_1+\mathbf{p}_2)/2$, $\mathbf{p}'=(\mathbf{p}_3+\mathbf{p}_4)/2$, $\mathbf{q}=(\mathbf{p}_1-\mathbf{p}_2)=(\mathbf{p}_3-\mathbf{p}_4)$, and finally obtain
\begin{equation}\label{incoh_sca_eq5}
\begin{split}
&I_{\text{ich}}(\mathbf{r})=n_0\sum_{mpm'p'}\iiiint d\mathbf{p}d\mathbf{p}'d\mathbf{q}d\mathbf{s}[n_0h_2(\mathbf{s})+\delta(\mathbf{s})]\\&\cdot T_{1p}T_{1p'}^*C_{mp}C_{m'p'}^* \mathbf{N}^{(3)}_{m1p}(\mathbf{p}'+\mathbf{q}/2)\mathbf{N}^{(3)*}_{m1p}(\mathbf{p}'-\mathbf{q}/2)\\&\cdot\exp[i(\mathbf{p}-\mathbf{p}')\cdot\mathbf{s}]\exp(i\mathbf{q}\cdot\mathbf{r})\langle \mathbf{E}(\mathbf{p}+\mathbf{q}/2)\mathbf{E}^*(\mathbf{p}-\mathbf{q}/2)\rangle.
\end{split}
\end{equation}
Integrating over $\mathbf{s}$ and using the Fourier representation of pair correlation function $h_2(\mathbf{s})$ as
\begin{equation}\label{Hq_eq}
H(\mathbf{p}'-\mathbf{p})=\frac{1}{(2\pi)^3}\int d\mathbf{s}h_2(s)\exp{[-i(\mathbf{p}'-\mathbf{p})\cdot\mathbf{s}]}.
\end{equation}

Therefore we are able to obtain
\begin{equation}\label{incoh_sca_eq6}
\begin{split}
&I_{\text{ich}}(\mathbf{r})=n_0\sum_{mpm'p'}\iiint d\mathbf{p}d\mathbf{p}'d\mathbf{q}[n_0(2\pi)^3H(\mathbf{p}'-\mathbf{p})+1] \\&\cdot T_{1p}T_{1p'}^*C_{mp}C_{m'p'}^* \mathbf{N}^{(3)}_{m1p}(\mathbf{p}'+\mathbf{q}/2)\mathbf{N}^{(3)*}_{m1p}(\mathbf{p}'-\mathbf{q}/2)\\&\cdot\langle \mathbf{E}(\mathbf{p}+\mathbf{q}/2)\mathbf{E}^*(\mathbf{p}-\mathbf{q}/2)\rangle\exp(i\mathbf{q}\cdot\mathbf{r}).
\end{split}
\end{equation}

Furthermore, we take \textit{on-shell} approximation for total intensity, i.e., $\langle \mathbf{E}(\mathbf{p}+\mathbf{q}/2)\mathbf{E}^*(\mathbf{p}-\mathbf{q}/2)\rangle$ is concentrated in a momentum shell at $p=K$, where $K$ is the effective propagation constant calculated before \cite{lagendijk1996resonant,vynckPRA2016}. This approximation is valid when scatterers are in the far field of each other and each scattering event occurs in the far field of each other. This condition also requires $\mathbf{q}\rightarrow 0$, meaning no \textit{off-shell} wave components enter into the total intensity. In this way, $\langle \mathbf{E}(\mathbf{p}+\mathbf{q}/2)\mathbf{E}^*(\mathbf{p}-\mathbf{q}/2)\rangle\approx \langle \mathbf{E}(\mathbf{p})\mathbf{E}^*(\mathbf{p})\rangle\delta(\mathbf{p}-\hat{\mathbf{p}}K)$. In the present study, the mean normalized distance between each two scatterers can be estimated to be $kd=2\pi a \sqrt[3]{4\pi/(3f_v)}/\lambda=2.42$ for the largest density case of $f_v=0.25$. Therefore we can assume that the far-field and on-shell approximations are applicable. In this circumstance, we carry out the integral involving $\mathbf{q}$ as
\begin{equation}\label{ftrans_vswf}
\begin{split}
&\iiint d\mathbf{q}d\mathbf{r}_1d\mathbf{r}_2\mathbf{N}^{(3)}_{m1p}(\mathbf{r}_1)\mathbf{N}^{(3)*}_{m1p}(\mathbf{r}_2)\langle\mathbf{E}(\mathbf{p}+\mathbf{q}/2)\mathbf{E}^*(\mathbf{p}-\mathbf{q}/2)\rangle\\&\cdot\exp(i\mathbf{q}\cdot\mathbf{r})\exp(-i(\mathbf{p}'+\mathbf{q}/2)\mathbf{r}_1)\exp(i(\mathbf{p}'-\mathbf{q}/2)\mathbf{r}_2)\\&\approx
\int d\mathbf{s}\mathbf{N}^{(3)}_{m1p}(\mathbf{r}+\mathbf{s}/2)\mathbf{N}^{(3)*}_{m1p}(\mathbf{r}-\mathbf{s}/2)\exp(-i\mathbf{p}'\cdot \mathbf{s})\\&\cdot\langle\mathbf{E}(\hat{\mathbf{p}}K)\mathbf{E}^*(\hat{\mathbf{p}}K)\rangle,
\end{split}
\end{equation}
where the definition of Fourier transform for VSWFs is used. Substituting Eq.(\ref{ftrans_vswf}) into Eq.(\ref{incoh_sca_eq6}), we have
\begin{equation}\label{incoh_sca_eq7}
\begin{split}
&I_{\text{ich}}(\mathbf{r})=n_0\sum_{mpm'p'}\iiint d\hat{\mathbf{p}}d\mathbf{p}'d\mathbf{s}[n_0(2\pi)^3H(\mathbf{p}'-\mathbf{p})+1] \\&\cdot T_{1p}T_{1p'}^*C_{mp}C_{m'p'}^* \mathbf{N}^{(3)}_{m1p}(\mathbf{r}+\mathbf{s}/2)\mathbf{N}^{(3)*}_{m1p}(\mathbf{r}-\mathbf{s}/2)\\&\cdot\exp(-i\mathbf{p}'\cdot \mathbf{s})\langle\mathbf{E}(\hat{\mathbf{p}}K)\mathbf{E}^*(\hat{\mathbf{p}}K)\rangle.
\end{split}
\end{equation}
By this stage, the physical significance of Eq.(\ref{incoh_sca_eq7}) is obvious. It describes that incoherent intensity arises from the process that total intensity propagating along $\hat{\mathbf{p}}$ is scattered into the direction $\hat{\mathbf{p}}'$, and the total incoherent intensity should be integrated over all possible incident and scattering directions. This is the process depicted by conventional RTE \cite{mishchenko2006multiple}. Therefore, the quantity in the integral is indeed the different scattering cross section of an individual scattering event, which is given by
\begin{equation}\label{pf_qca}
\begin{split}
&\frac{d\sigma_\text{s}'}{d\varOmega_\text{s}}=n_0\sum_{mpm'p'}\int d\mathbf{s}[n_0(2\pi)^3H(\mathbf{p}'-\mathbf{p})+1] \\&\cdot T_{1p}T_{1p'}^*C_{mp}C_{m'p'}^* \mathbf{N}^{(3)}_{m1p}(\mathbf{r}+\mathbf{s}/2)\mathbf{N}^{(3)*}_{m1p}(\mathbf{r}-\mathbf{s}/2)\\&\cdot\exp(-i\mathbf{p}'\cdot \mathbf{s}),
\end{split}
\end{equation}
where $\varOmega_\text{s}$ indicates the scattering solid angle defined as the angle between incident direction $\mathbf{p}$ and scattering direction $\mathbf{p}'$. Since we have assumed far-field scattering, above equation can be calculated by utilizing the asymptotic property for VSWFs in the far field, which is listed in Appendix \ref{far-field_appendix}. The integral over $\mathbf{s}$ results in a requirement that the scattering momentum $\mathbf{p}'$ should be equal to $k\hat{\mathbf{r}}$, which is consistent with the far-field behavior of VSWFs, implying that only the mode with momentum $p'=k$ can propagate into the far field. The momentum mismatch between $p=K$ and $p'=k$ is actually compensated by the term involving the Fourier transform of the pair correlation function $H(\mathbf{p}'-\mathbf{p})$. After some manipulations we obtain 
\begin{equation}\label{pf_qca2}
\begin{split}
&\frac{d\sigma_\text{s}'}{d\theta_\text{s}}=\frac{9n_0}{4k^2}[1+n_0(2\pi)^3H(\mathbf{p}'-\mathbf{p})]\\
&\times\Big[|a_1C_{21}\pi_n(\cos\theta_\text{s})+b_1C_{11}\tau_n(\cos\theta_\text{s})|^2\\
&+|b_1C_{11}\pi_n(\cos\theta_\text{s})+a_1C_{21}\tau_n(\cos\theta_\text{s})|^2\Big],
\end{split}
\end{equation}
where $\theta_\text{s}$ is the polar scattering angle, and the dependency on azimuth angle is integrated out. The functions $\tau_n(\cos\theta_\text{s})$ and $\pi_n(\cos\theta_\text{s})$ are defined in Appendix \ref{far-field_appendix}. 

By now the only undetermined object is the Fourier transform of pair correlation function, $H(\mathbf{p}'-\mathbf{p})$, where the absolute value of argument is $|\mathbf{p}'-\mathbf{p}|=\sqrt{K^2+k^2-2Kk\cos\theta_s}$. To this end, here we regard that all silicon particles are randomly distributed and the only restriction is that they do not overlap or penetrate each other. In this assumption, the Percus-Yevick approximation for the PDF of hard spheres is used, since this model is capable to reproduce the position relations between pairs of spherical particle analytically with high accurateness \cite{tsang2004scattering2}. It is given by Refs.\cite{wertheimPRL1963,tsang2004scattering2} as
\begin{equation}
H(\mathbf{q})=\frac{F(q)}{1-n_0(2\pi)^3F(q)},
\end{equation}
where
\begin{equation}
\begin{split}
F(q)=&24f_v[\frac{\alpha+\beta+\delta}{u^2}\cos u-\frac{\alpha+2\beta+4\delta}{u^3}\sin u\\&-2\frac{\beta+6\delta}{u^4}\cos u+\frac{2\beta}{u^4}+\frac{24\delta}{u^6}(\cos u-1)]
\end{split}
\end{equation}
with $q=|\mathbf{q}|$, $u=4qa$, $\alpha=(1+2f_v)^2/(1-f_v)^4$, $\beta=-6f_v(1+f_v/2)^2/(1-f_v)^4$, $\delta=f_v(1+2f_v)^2/[2(1-f_v)^2]$. 
\begin{figure}[htbp]
	\centering
	\includegraphics[width=0.8\linewidth]{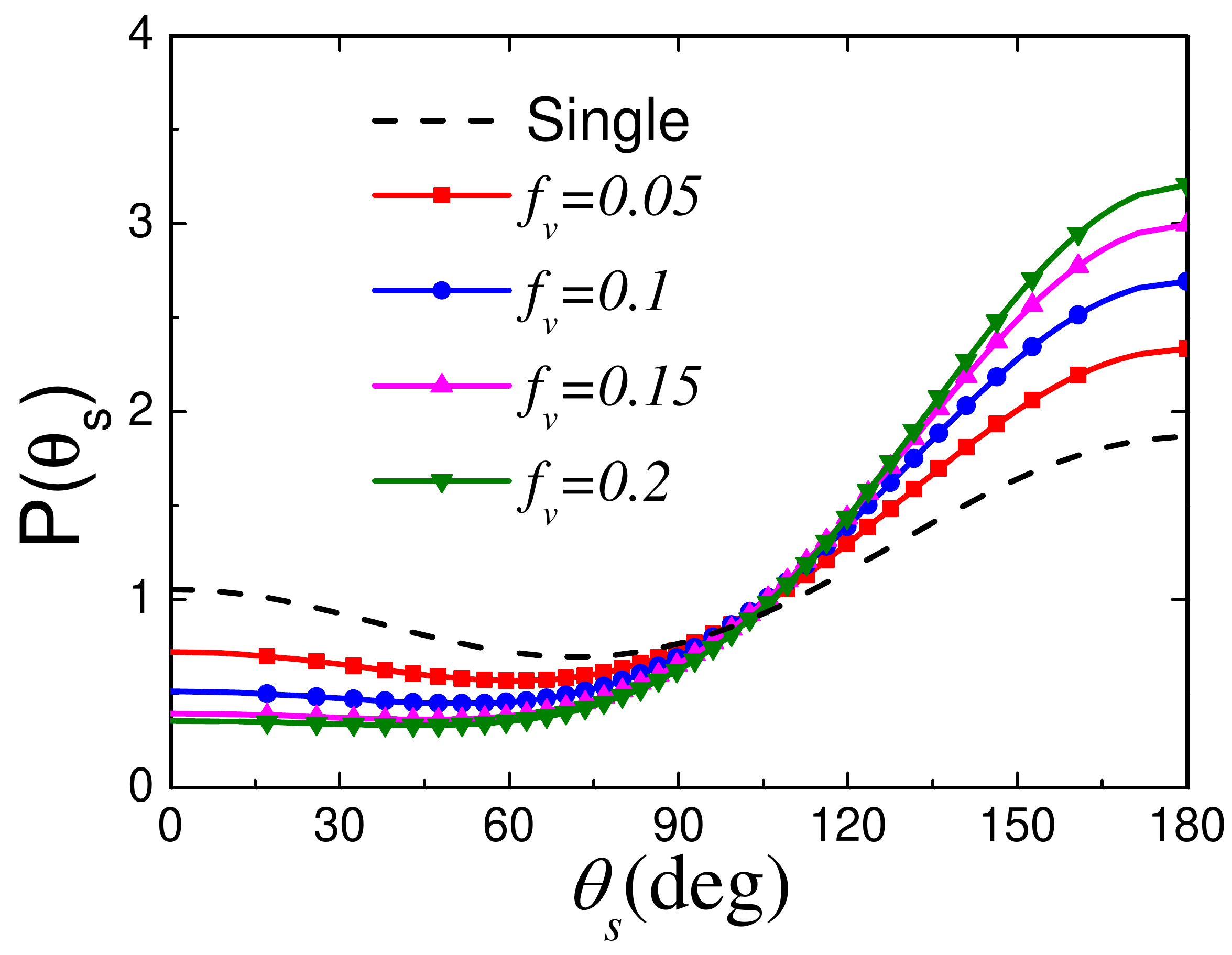}
	\caption{QCA-calculated phase function $P(\theta_\text{s})$ as a function of scattering angle $\theta_\text{s}$ for the random medium containing Si nanoparticles with radius $a=230\mathrm{nm}$ with different volume fractions $f_v$ at wavelength $\lambda=1530\mathrm{nm}$. The interacting potential between particles is the hard-sphere type calculated through Percus-Yevick model \cite{wertheimPRL1963}.}
	\label{figpf}
\end{figure}

The obtained phase functions under QCA for different volume fractions are provided in Fig. \ref{figpf}. Here the upper limit of volume fraction is chosen to be $f_v=0.25$ since higher volume fraction of particles would lead to the breakdown of QCA, because particle correlations involving three or more particles are more complicated than what QCA predicts. Specifically, the phase function is computed through normalizing differential scattering cross section as \cite{mishchenko2006multiple}
\begin{equation}
P(\theta_\text{s})=\frac{\frac{d\sigma_\text{s}'}{d\theta_\text{s}}}{\frac{1}{2}\int_{0}^{\pi}\frac{d\sigma_\text{s}'}{d\theta_\text{s}}\sin\theta_\text{s}d\theta_\text{s}}.
\end{equation}
Fig.\ref{figpf} demonstrates that by increasing the particle concentration, the forward scattering is reduced and backscattering is enhanced, and the effect of $f_v$ is more pronounced on backscattering than forward scattering. Therefore, we have achieved much stronger backscattering phase functions than that in the single scattering case. This is one of the main results of the present paper.

To understand the underlying mechanism on this enhancement for backscattering, it is worth giving a brief exploration on the dependent scattering effects, including the modification of electric and magnetic dipole excitations and far-field interference effect, both induced and influenced by the structural correlations. 

\begin{figure}[htbp]
	\centering
	\includegraphics[width=0.8\linewidth]{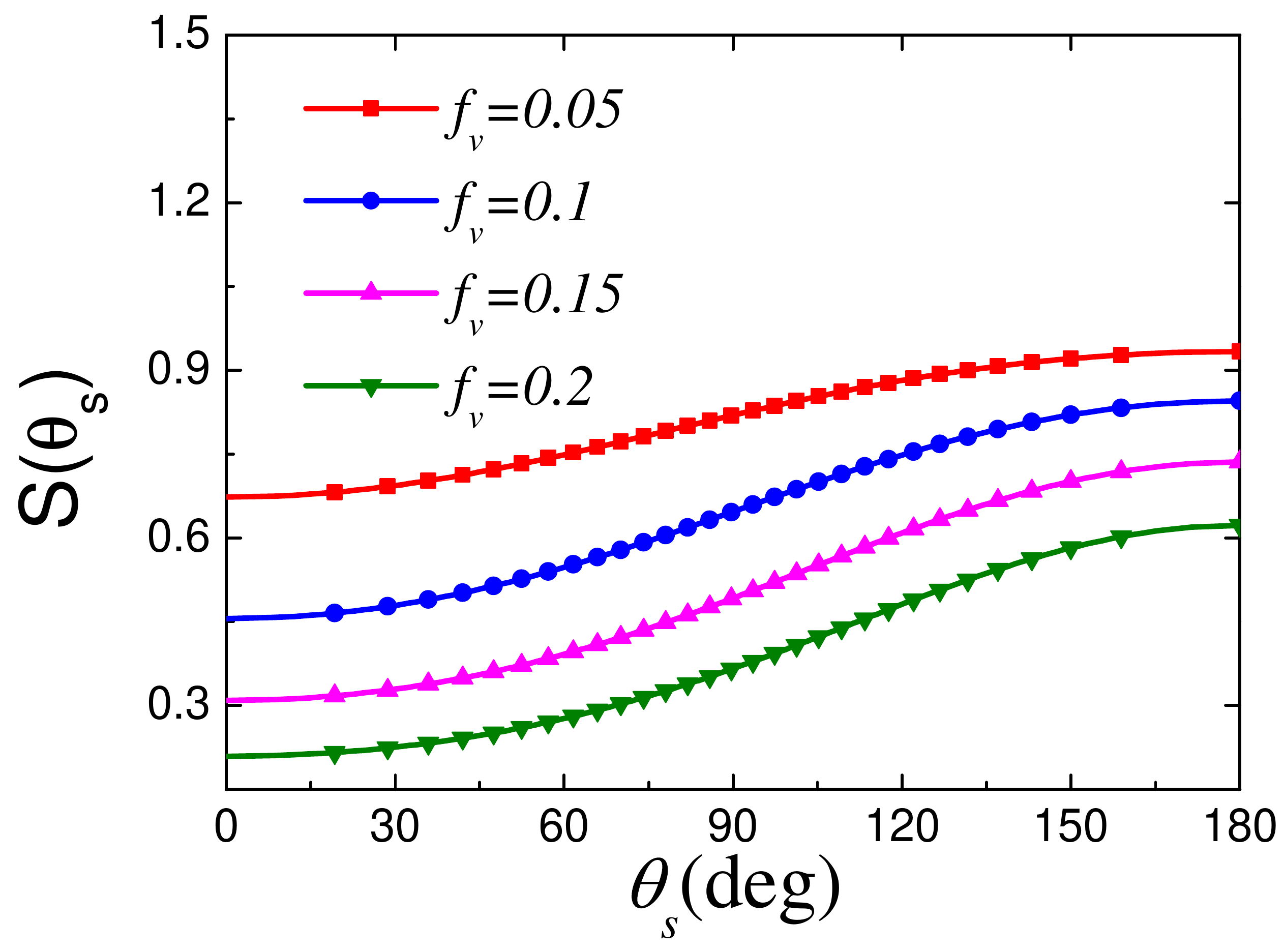}
	\caption{Structure factor $S$ as a function of scattering angle $\theta_s$ for the random medium containing Si nanoparticles with radius $a=230\mathrm{nm}$ with different volume fractions $f_v$ at wavelength $\lambda=1530\mathrm{nm}$. The interacting potential between particles is the hard-sphere type calculated through Percus-Yevick model \cite{wertheimPRL1963}.}
	\label{figsq}
\end{figure}

In fact, a close scrutiny of Eq.(\ref{pf_qca2}) provides the clear physical significance of the differential scattering cross section. The structural correlations among particles, i.e., $g_2(r)$, enter into Eq.(\ref{pf_qca}) and affects the dependent scattering mechanism in two ways. The first is contained in the fluctuational component of structure factor defined as $S(\mathbf{q})=1+n_0(2\pi)^3H(\mathbf{p}'-\mathbf{p})$ where $\mathbf{q}=\mathbf{p}'-\mathbf{p}$. This quantity is widely used by many authors as the first order dependent-scattering correction to $d\sigma_s/d\theta_s$ of single particle scattering, for instance, Refs.\cite{fradenPRL1990,mishchenkoJQSRT1994,rojasochoaPRL2004,yamadaJHT1986,conleyPRL2014}, which describes the far-field interference between first-order scattered waves of different particles, also named as the interference approximation (ITA) by some authors \cite{dickJOSAA1999}. The only difference in $S(\mathbf{q})$ between QCA and ITA is that the former takes the propagation constant of effective excitation field $\mathbf{K}$ impinging on the particles into account, rather than the bare value of $\mathbf{k}$. In Fig.\ref{figsq}, we give $S(\mathbf{q})$ for different volume fractions as a function of scattering angle $\theta_s$, in which $q=\sqrt{K^2+k^2-2Kk\cos\theta_s}$ is implicitly used. Results show that the effect of $S(\theta_s)$ is that it reduces forward incoherent scattering intensity more significantly than backscattering intensity. The back/forward scattering contrast grows with the increasing of volume fraction, or the degree of structural correlations. 

\begin{figure}[htbp]
	\centering
	\includegraphics[width=0.8\linewidth]{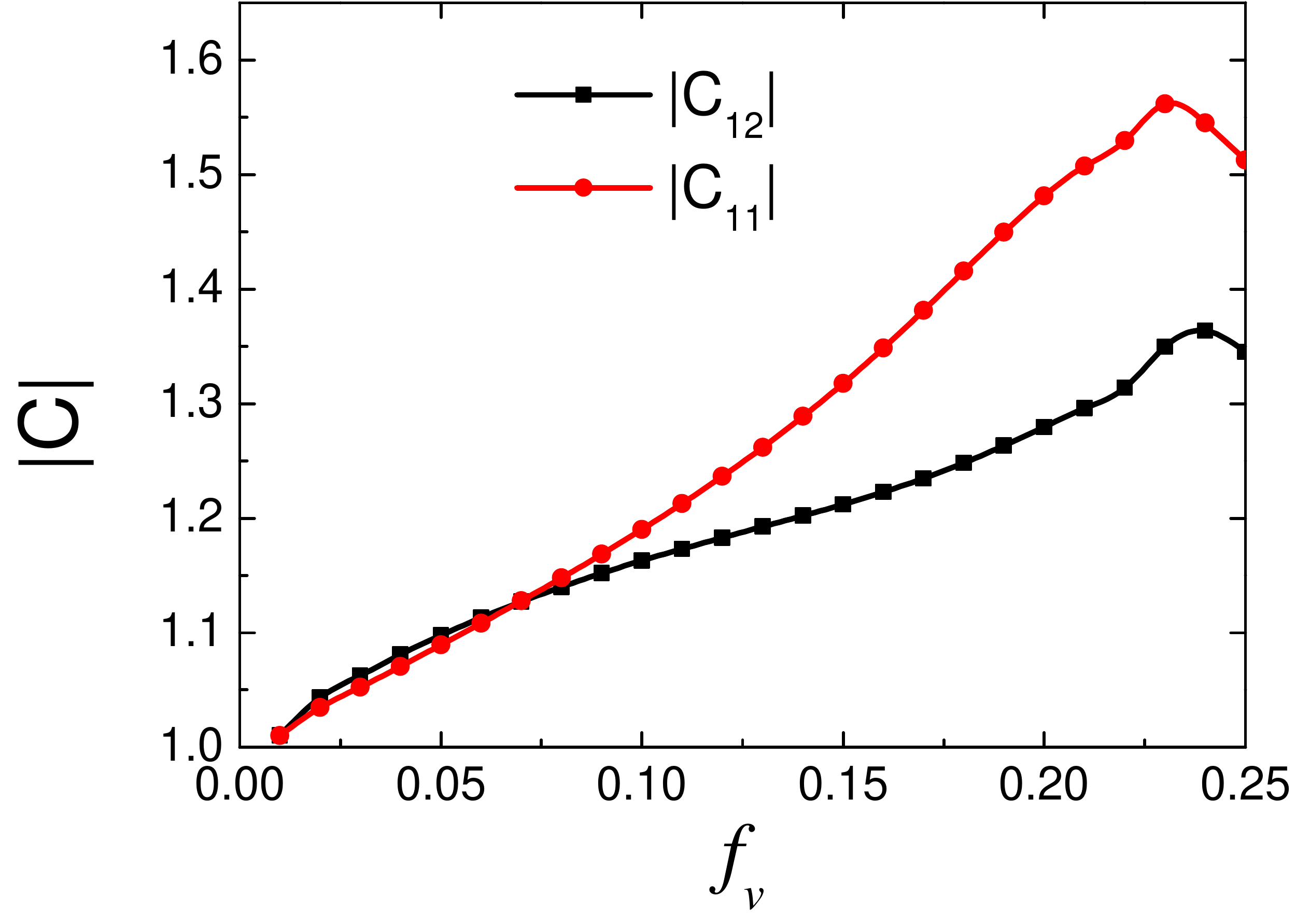}
	\caption{ Variation of effective exciting field amplitudes $C_{12}$ (for electric dipole) and $C_{11}$ (for magnetic dipole) with particle volume fraction $f_v$.}\label{xexm}
	\label{figxexm}
\end{figure}

The second role of the structural correlations is that they affect the effective exciting field amplitudes for electric and magnetic dipoles, $C_{12}$ and $C_{11}$ according to Eq.(\ref{aprox2}), which are both equal to 1 under ISA as well as ITA.  In Fig.\ref{xexm}, we have calculated the absolute values of $C_{11}$ and $C_{12}$ to demonstrate how structural correlations induce an dependent scattering effect on the dipole excitation under different volume fractions. It is found that the absolute values of $C_{12}$ and $C_{11}$ are substantially larger than 1, which indicates that dependent scattering mechanism gives rise to an enhancement in mode amplitudes for dual-dipolar particles compared to the free-space plane-wave illumination. Both $C_{12}$ and $C_{11}$ grow with volume fraction until $f_v\gtrsim0.23$, where they start to decrease. This suggests that dependent scattering mechanism initially enhances the electromagnetic excitation of particles at moderate concentrations and when the volume fraction continues to rise a reduction occurs due to the ``screening effects'', in which individual particle ``witnesses'' its surrounding medium as a high-index-of-refraction effective medium, rather than the bare background medium (vacuum in the present case), leading to a reduction in index contrast and thus scattering strength \cite{sheng2006introduction}. This is actually a renormalization of wave propagation in random media, which has also been considered by many authors but with rather different approaches, for instance, Ref.\cite{busch1995PRL,busch1996PRB,sheng2006introduction,Liew2011,naraghiOL2015}. Moreover, the amplitude of $C_{11}$ is significantly larger than $C_{12}$, implying that in the current case, dependent scattering enhances magnetic dipole excitation more efficiently.

\begin{figure}[htbp]
	\centering
	\includegraphics[width=0.8\linewidth]{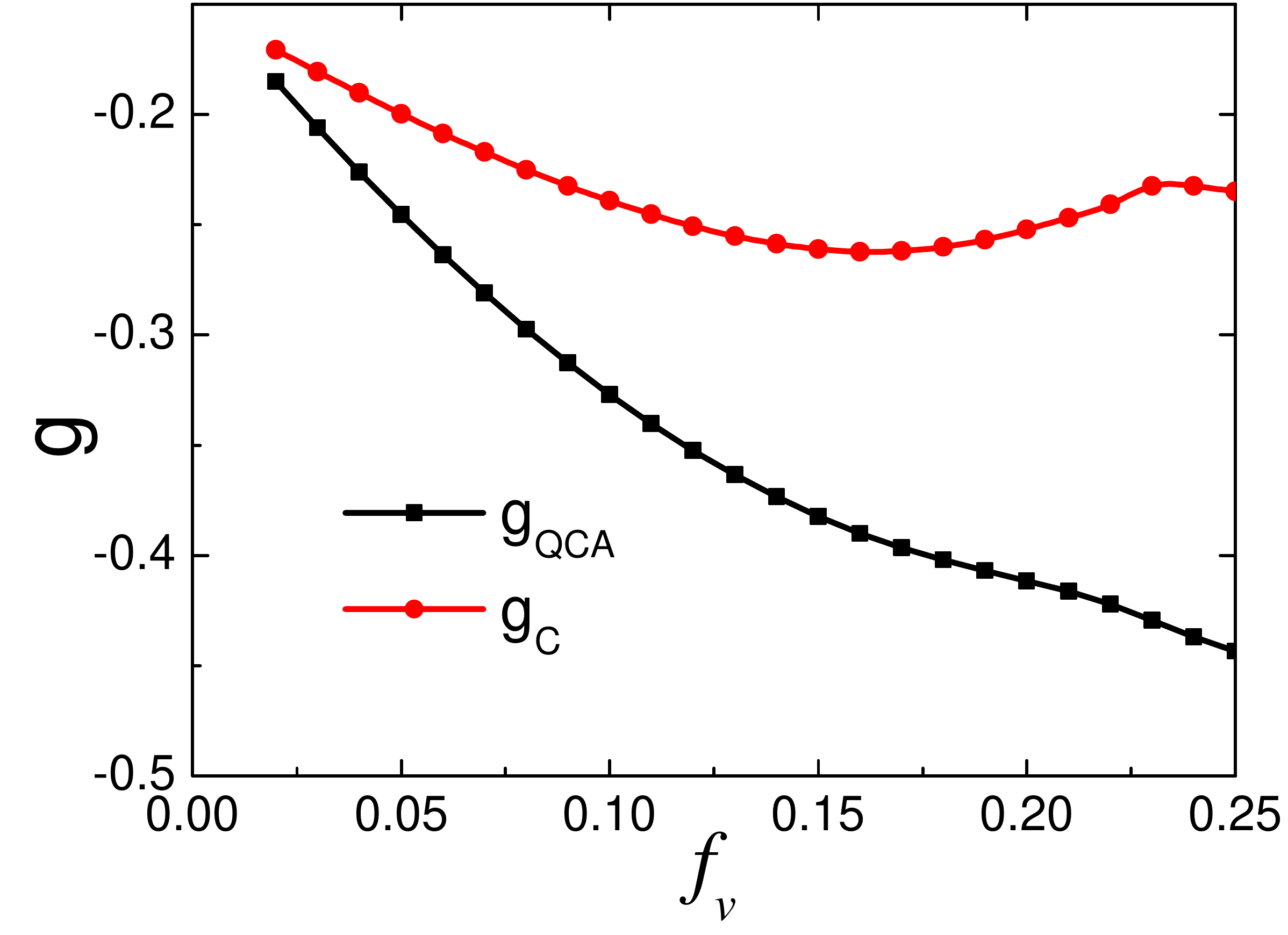}
	\caption{QCA-calculated asymmetry factor $g_\text{QCA}$ for a random medium containing Si nanoparticle with radius $a=230\mathrm{nm}$ with different volume fractions $f_v$  at wavelength $\lambda=1530\mathrm{nm}$, compared with the partial asymmetry factor $g_\text{C}$ that only considers the effect of the modification of the exciting field induced by the structural correlations.}
	\label{figgqca}
\end{figure}

According to Eq.(\ref{pf_qca}), knowing that $\pi_1(1)=\tau_1(1)=\pi_1(-1)=-\tau_1(-1)=1$, the differential scattering cross section in forward and backward scattering direction can be estimated as 
\begin{equation}\label{pf_qca1}
\begin{split}
\frac{d\sigma_\text{s}'}{d\theta_\text{s}}|_{\theta_\text{s}=0^\text{o}}\sim|a_1C_{12}+b_1C_{11}|^2,
\end{split}
\end{equation}
\begin{equation}\label{pf_qca3}
\begin{split}
\frac{d\sigma_\text{s}'}{d\theta_\text{s}}|_{\theta_\text{s}=180^\text{o}}\sim|a_1C_{12}-b_1C_{11}|^2,
\end{split}
\end{equation}
if the prefactor containing $S(\mathbf{q})$ is not taken into account.
Thus it is straightforward to obtain \cite{Gomez-MedinaPRA2012}
\begin{equation}\label{gc}
g_{\text{QCA}}=\frac{1}{2}\frac{\frac{d\sigma_\text{s}'}{d\theta_\text{s}}|_{\theta_\text{s}=0^\text{o}}-\frac{d\sigma_\text{s}'}{d\theta_\text{s}}|_{\theta_\text{s}=180^\text{o}}}{\frac{d\sigma_\text{s}'}{d\theta_\text{s}}|_{\theta_\text{s}=0^\text{o}}+\frac{d\sigma_\text{s}'}{d\theta_\text{s}}|_{\theta_\text{s}=180^\text{o}}}\sim\frac{\mathrm{Re}(a_1C_{12}b_1^*C_{11}^*)}{|a_1C_{12}|^2+|b_1C_{11}|^2}.
\end{equation}
We represent the quantity in the right hand side of Eq. (\ref{gc}) as $g_{\text{C}}$, denoting solely the contribution from correlation induced dependent-scattering effect on the exciting field. The results of  asymmetry factor from full QCA, $g_{\text{QCA}}$, along with $g_{\text{C}}$ are compared in Fig.\ref{figgqca}, in which this partial contribution on the negative asymmetry factor is observed to be substantial, more negative than that of single particles. Consequently we have unequivocally demonstrated the second role of the structural correlations, i.e., modification of the exciting field for the dipolar modes, which also leads to an enhancement in backscattering. This role, actually, is rarely noticed or explicitly demonstrated by previous studies.
\begin{figure}[htbp]
	\centering
	\includegraphics[width=0.8\linewidth]{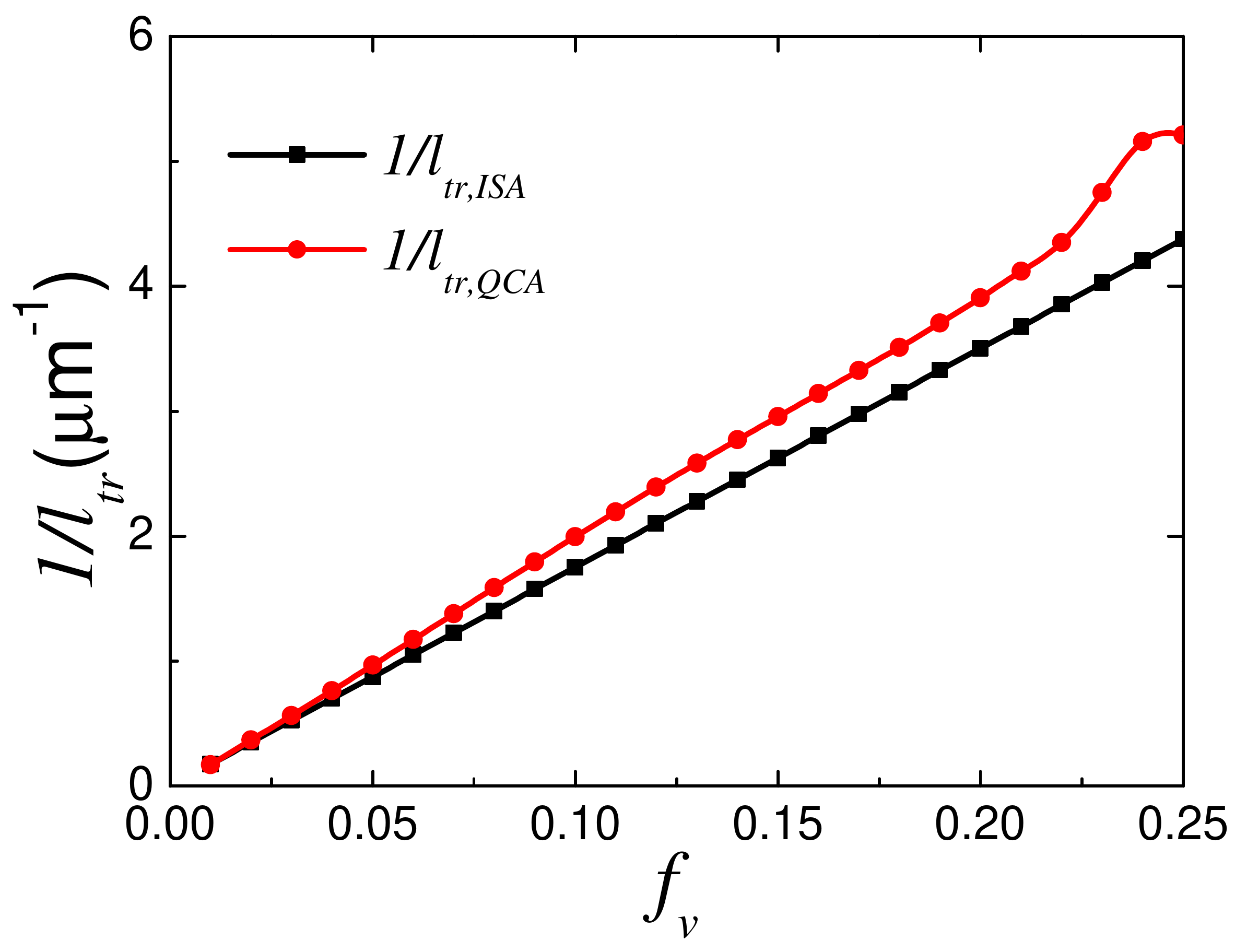}
	\caption{The inverse of transport mean free path $1/l_\text{tr}$ for a random medium containing Si nanoparticle with radius $a=230\mathrm{nm}$ with different volume fractions $f_v$ calculated by QCA, where the operating wavelength is $\lambda=1530\mathrm{nm}$, and the result of ISA is also plotted for comparison.}
	\label{figltr}
\end{figure}

Finally, we calculate the transport mean free path under QCA with comparison with that under ISA as a function of volume fraction. To show the linear dependence of scattering strength predicted by ISA, the inverse of transport mean free path $1/l_\text{tr}$ is used in Fig.\ref{figltr}. In the studied range of $f_v$, the inverse of transport mean free path under QCA is surprisingly higher that of ISA, implying a much lower $l_\text{tr}$ and then lower light conductance \cite{lagendijk1996resonant,sheng2006introduction}. The highest value of $1/l_\text{tr,QCA}$ appears at $f_v\sim0.23$, leading to the Ioffe-Regel parameter $Kl_\text{tr,QCA}\sim 1$, which indicates that strong localization behavior may occur. This result is surprising because in conventional densely-packed particle systems, the structural correlations strongly reduce the scattering strength, leading to a giant increase of the transport mean free path compared to the value predicted by ISA \cite{fradenPRL1990}. It is interesting to examine the possibility of localization in more details, which will be our further work.

\section{Conclusions}
In this paper we have accomplished the design and theoretical analysis of a random medium with a strongly negative scattering asymmetry factor for multiple light scattering in the near-infrared. Based on a multipole expansion of the FLEs and QCA, we have rigorously derived analytical expressions for the effective propagation constant for a random system containing dual-dipolar particles. Moreover, in terms of intensity transport, we derive a Bethe-Salpeter-type equation for this system. By applying far-field and on-shell approximations as well as Fourier transform technique, we have finally obtained the scattering phase function, which is one of the main contributions of the present paper. Although the present formulas are deduced only for dual-dipolar particles, our theory can be naturally extended to take high-order multipoles into account.
	
Following from our theoretical contribution, by utilizing structural correlations among particles and the second Kerker condition, we show that for a random medium composed of randomly distributed silicon nanoparticles, the asymmetry factor can reach nearly $g\sim-0.5$. The structural correlations are of the hard-sphere type and we find that as concentration of particles rises, the backscattering is also enhanced. This leads to a strong reduction in the transport mean free path, even lower than the value calculated from ISA, resulting in a potential for strong localization to occur. We further reveal that the strongly backscattering phase function is a result of the dependent scattering effects, including the modification of electric and magnetic dipole excitations and far-field interference effect, which are both induced and influenced by the structural correlations.

To sum up, in the theoretical aspect, the present study establishes a basis for analyzing light propagation, including field and intensity, in disordered media with multipole Mie modes. In the application aspect, it paves a new way to manipulate light scattering and spawns new possibilities such as imaging, photovoltaics and radiative cooling through random media. For instance, a strongly backscattering feature is promising for radiative cooling coatings \cite{baoSEMSC2017,zhaiScience2017}, for which it is necessary to efficiently reflect incident solar power concentrating in visible and near-infrared.
\section*{Acknowledgments}
This work is supported by the National Natural Science Foundation of China (Grant Nos. 51636004 and 51476097), Shanghai Key Fundamental Research Grant (Grant No. 16JC1403200) as well as the the Foundation for Innovative Research Groups of the National Natural Science Foundation of China (Grant No. 51521004).
\appendix
\section{Definition of ensemble average}\label{en_avg_def}
Generally, an ensemble average of a physical quantity should be carried out with respect to \textit{all} possible states of the system \cite{laxRMP1951,tsang2004scattering2,mishchenko2006multiple}. In the present random medium consisting of $N$ identical, homogeneous and isotropic nanoparticles, the only varying states of particles are their positions, i.e., $\mathbf{r}_j$, where $j=1, 2,...,N$. Therefore, the ensemble average over the whole random medium for a physical quantity $Q(\mathbf{r}_1,\mathbf{r}_2,...,\mathbf{r}_j,...,\mathbf{r}_N)$, which is a function of particle positions, is calculated as \cite{laxRMP1951,tsang2004scattering2}
\begin{equation}
\begin{split}
\langle Q\rangle=\int& Q(\mathbf{r}_1,\mathbf{r}_2,...,\mathbf{r}_j,...,\mathbf{r}_N)d\mathbf{r}_1d\mathbf{r}_2...d\mathbf{r}_j...d\mathbf{r}_N\\&\cdot p(\mathbf{r}_1,\mathbf{r}_2,...,\mathbf{r}_j,...,\mathbf{r}_N)
\end{split}
\end{equation}	
where $p(\mathbf{r}_1,\mathbf{r}_2,...,\mathbf{r}_j,...,\mathbf{r}_N)$ is the joint probability density function of the particle distribution $\mathbf{r}_1,\mathbf{r}_2,...,\mathbf{r}_j,...,\mathbf{r}_N$. If we fix some particle $\mathbf{r}_j$, the ensemble average over other $N-1$ particles is given by
\begin{equation}
\begin{split}
\langle Q\rangle_j=\int& Q(\mathbf{r}_1,\mathbf{r}_2,...,\mathbf{r}_i,...,\mathbf{r}_j,...,\mathbf{r}_N)d\mathbf{r}_1d\mathbf{r}_2...d\mathbf{r}_i...d\mathbf{r}_N\\&\cdot p(\mathbf{r}_1,\mathbf{r}_2,...,\mathbf{r}_i,...,\mathbf{r}_j,...,\mathbf{r}_N)
\end{split}
\end{equation}
where $i\ne j$.	
Similarly, if we fix two particles $\mathbf{r}_i$ and $\mathbf{r}_j$, the ensemble average is expressed as
\begin{equation}
\begin{split}
\langle Q\rangle_{ij}=\int& Q(\mathbf{r}_1,\mathbf{r}_2,...,\mathbf{r}_i,...,\mathbf{r}_j,...,\mathbf{r}_l,...\mathbf{r}_N)d\mathbf{r}_1d\mathbf{r}_2...d\mathbf{r}_l...d\mathbf{r}_N\\&\cdot p(\mathbf{r}_1,\mathbf{r}_2,...,\mathbf{r}_i,...,\mathbf{r}_j,...,\mathbf{r}_l,...\mathbf{r}_N)
\end{split}
\end{equation}
where $i\ne j$, $l\ne j$ and $l\ne i$.
The relation between $\langle Q\rangle_{ij}$ and $\langle Q\rangle_{j}$ can be derived as
\begin{equation}
\langle Q\rangle_j=\int\langle Q\rangle_{ij}p(\mathbf{r}_i|\mathbf{r}_j)d\mathbf{r}_i
\end{equation}
where $p(\mathbf{r}_i|\mathbf{r}_j)$ is the conditional probability density function of $\mathbf{r}_i$ for a fixed $\mathbf{r}_j$. The pair distribution function, is related to $p(\mathbf{r}_i|\mathbf{r}_j)$ as \cite{tsang2004scattering2}
\begin{equation}
p(\mathbf{r}_i|\mathbf{r}_j)=\frac{g_2(\mathbf{r}_i|\mathbf{r}_j)}{V}\frac{N}{N-1}
\end{equation}
where $V$ is the volume occupied by the ensemble of particles. In the thermodynamic limit, $N\rightarrow \infty$, $p(\mathbf{r}_i|\mathbf{r}_j)\approx g_2(\mathbf{r}_i|\mathbf{r}_j)/V$.

\section{VSWFs and translation addition theorem}\label{vswf_appendix}
The regular VSWFs $\mathbf{N}^{(1)}_{mnp}(\mathbf{r})$ for $p=2$ (TE mode) and $p=1$ (TM mode) are defined as \cite{mackowskiJOSAA1996,mackowskiJQSRT2013,tsang2000scattering1,bohrenandhuffman,hulst1957}
\begin{equation}
\mathbf{N}^{(1)}_{mn2}(\mathbf{r})=\sqrt{\frac{(2n+1)(n-m)!}{4\pi n(n+1)(n+m)!}}\nabla\times(\mathbf{r}\psi_{mn}^{(1)}(\mathbf{r})),
\end{equation}
\begin{equation}
\mathbf{N}^{(1)}_{mn1}(\mathbf{r})=\frac{1}{k}\nabla\times\mathbf{N}^{(1)}_{mn2}(\mathbf{r})
\end{equation}
where $k=\omega/c$ is the wave number in free space and $\omega$ is the angular frequency of the electromagnetic wave. $\psi_{mn}^{(1)}(\mathbf{r})$ is regular (type-1) scalar wave function defined as
\begin{equation}
\psi_{mn}^{(1)}(\mathbf{r})=j_n(kr)Y_n^m(\theta,\phi),
\end{equation}
where $j_n(kr)$ is the spherical Bessel function and $Y_n^m(\theta,\phi)$ is spherical harmonics defined as
\begin{equation}
Y_n^m(\theta,\phi)=P_n^m(\cos\theta)\exp(im\phi),
\end{equation}
where we use the convention of quantum mechanics, and $P_n^m(\cos\theta)$ is associated Legendre polynomials.

The outgoing (type-3) VSWFs have can be similarly defined by replacing above spherical Bessel functions with Hankel functions of the first kind $h_n(kr)$.

The translation addition theorem of VSWFs, which transforms the VSWFs centered in $\mathbf{r}_i$ into those centered in $\mathbf{r}_j$, is given by
\begin{equation}
\mathbf{N}^{(3)}_{\mu\nu q}(\mathbf{r}-\mathbf{r}_i)=\sum_{\mu\nu q}A_{mnp\mu\nu q}^{(3)}(\mathbf{r}_i-\mathbf{r}_j)\mathbf{N}^{(1)}_{mnp}(\mathbf{r}-\mathbf{r}_j),
\end{equation}
which is valid for $|\mathbf{r}_i-\mathbf{r}_j|>|\mathbf{r}-\mathbf{r}_j|$, and therefore should be used in the vicinity of $\mathbf{r}_j$. The coefficient $A_{\mu qmp}^{(3)}$ is generally given by \cite{tsang2004scattering2}
\begin{equation}\label{translation_coef3}
\begin{split}
&A_{mn1\mu\nu 1}^{(3)}(\mathbf{r})=A_{mn2\mu\nu 2}^{(3)}(\mathbf{r})=\frac{\gamma_{\mu\nu}}{\gamma_{mn}}(-1)^{m}\\&\cdot\sum_{l}a(\mu,\nu|-m,n|l)a(\nu,n,l)h_l(kr)Y_{l}^{\mu-m}(\theta,\phi),
\end{split}
\end{equation}
\begin{equation}\label{translation_coef4}
\begin{split}
&A_{mn1\mu\nu 2}^{(3)}(\mathbf{r})=A_{mn2\mu \nu 1}^{(3)}(\mathbf{r})=\frac{\gamma_{\mu\nu}}{\gamma_{mn}}(-1)^{m+1}\\&\sum_{l}a(\mu,\nu|-m,n|l,l-1)b(\nu,n,l)h_l(kr)Y_{l}^{\mu-m}(\theta,\phi),
\end{split}
\end{equation}
where $\gamma_{mn}$ is defined as
\begin{equation}
\gamma_{mn}=\sqrt{\frac{(2n+1)(n-m)!}{4\pi n(n+1)(n+m)!}}.
\end{equation}
The coefficients $a(\mu,\nu|-m,n|l)$ and $a(\mu,\nu|-m,n|l,l-1)$ are given by
\begin{equation}
\begin{split}
&a(\mu,\nu|-m,n|l)=(-1)^{\mu-m}\left( 2l+1 \right) \left( \begin{matrix}
\nu&	 n&		l\\
\mu&    -m&		\mu-m\\
\end{matrix} \right) \\&\cdot\left( \begin{matrix}
\nu&	n&		l\\
0&		0&		0\\
\end{matrix} \right)\Big[\frac{(\nu+\mu)!(n-m)!(l-\mu+m)!}{(\nu-\mu)!(n+m)!(l+\mu-m)!}\Big]^{1/2},
\end{split}
\end{equation}
\begin{equation}
\begin{split}
&a(\mu,\nu|-m,n|l,l-1)=(-1)^{\mu-m}\left( 2l+1 \right) \left( \begin{matrix}
\nu&	 n&		l\\
\mu&    -m&		\mu-m\\
\end{matrix} \right) \\&\cdot\left( \begin{matrix}
\nu&	n&		l-1\\
0&		0&		0\\
\end{matrix} \right)\Big[\frac{(\nu+\mu)!(n-m)!(l-\mu+m)!}{(\nu-\mu)!(n+m)!(l+\mu-m)!}\Big]^{1/2},
\end{split}
\end{equation}
in which the variables in the form $\left( \begin{matrix}
j_1&	j_2&		j_3\\
m_1&		m_2&	m_3\\
\end{matrix} \right)$ are Wigner-3$j$ symbols. They can be found in Ref. \cite{abramowitz1964handbook,tsang2004scattering2} and not shown in detail here. Other coefficients $a(\nu,n,l)$ and $b(\nu,n,l)$ are given as \cite{tsang2004scattering2}
\begin{equation}
\begin{split}
&a(\nu,n,l)=\frac{i^{n+l-\nu}}{2n(n+1)}\Big[2n(n+1)(2n+1)+(n+1)(\nu+n\\&-l)
(\nu+l-n+1)-n(\nu+n+l+2)(n+l-\nu+1)\Big],
\end{split}
\end{equation}
\begin{equation}
\begin{split}
&b(\nu,n,l)=-\frac{(2n+1)i^{n+l-\nu}}{2n(n+1)}\Big[(\nu+n+l+1)(n+l-\nu)\\\cdot&
(\nu+l-n)(\nu+n-l+1)\Big]^{1/2}.
\end{split}
\end{equation}
\section{Far-field approximation for outgoing VSWFs}\label{far-field_appendix}

For outgoing (type-3) VSWFs $\mathbf{N}^{(3)}_{mnp}(\mathbf{r}-\mathbf{r}_j)$ centered at $\mathbf{r}_j$, their far-field forms (when $r\gg r_j$) are given by \cite{mackowskiJOSAA1996,tsang2000scattering1,bohrenandhuffman}
\begin{equation}
\begin{split}
\mathbf{N}^{(3)}_{mn2}(\mathbf{r}-\mathbf{r}_j)&\approx i^{-n}\sqrt{\frac{(2n+1)(n-m)!}{4\pi n(n+1)(n+m)!}}\frac{\exp (kr)}{kr}\\&\cdot\exp (-\mathbf{k}_s\cdot\mathbf{r}_j)\mathbf{B}_{mn}(\theta,\phi),
\end{split}
\end{equation}
\begin{equation}
\begin{split}
\mathbf{N}^{(3)}_{mn1}(\mathbf{r}-\mathbf{r}_j)&\approx i^{-n}\sqrt{\frac{(2n+1)(n-m)!}{4\pi n(n+1)(n+m)!}}\frac{\exp(kr)}{kr}\\&\cdot\exp (-\mathbf{k}_s\cdot\mathbf{r}_j)\mathbf{C}_{mn}(\theta,\phi),
\end{split}
\end{equation}
where $\mathbf{B}_{mn}(\theta,\phi)$ and $\mathbf{C}_{mn}(\theta,\phi)$ are vector spherical harmonics. In the present calculation for spheres, only $m=\pm1$ are needed. In this condition,
\begin{equation}
\mathbf{B}_{1n}(\theta,\phi)=-[\hat{\bm{\theta}}\tau_n(\cos\theta)+\hat{\bm{\phi}}\pi_n(\cos\theta)]\exp(i\phi),
\end{equation}
\begin{equation}
\mathbf{B}_{-1n}(\theta,\phi)=\frac{1}{n(n+1)}[\hat{\bm{\theta}}\tau_n(\cos\theta)-\hat{\bm{\phi}}\pi_n(\cos\theta)]\exp(-i\phi),
\end{equation}
\begin{equation}
\mathbf{C}_{1n}(\theta,\phi)=-[\hat{\bm{\theta}}i\pi_n(\cos\theta)-\hat{\bm{\phi}}\tau_n(\cos\theta)]\exp(i\phi),
\end{equation}
\begin{equation}
\mathbf{C}_{-1n}(\theta,\phi)=-\frac{1}{n(n+1)}[\hat{\bm{\theta}}i\pi_n(\cos\theta)+\hat{\bm{\phi}}\tau_n(\cos\theta)]\exp(-i\phi),
\end{equation}
where $\tau_n$ and $\pi_n$ are functions defined as \cite{bohrenandhuffman}
\begin{equation}
\tau_n(\cos\theta)=-\frac{dP_n^1(\cos\theta)}{d\theta},
\end{equation}
\begin{equation}
\pi_n(\cos\theta)=-\frac{P_n^1(\cos\theta)}{\sin\theta}.
\end{equation}
\bibliography{neg_asym_factor_one_pr}

\end{document}